\documentclass[useAMS,usenatbib]{mn2e}

\usepackage{graphicx} 
\usepackage[usenames,dvipsnames]{color}
\usepackage{latexsym}
\usepackage{amssymb}
\usepackage{amsmath}
\usepackage{soul}

\newcommand{\aap}{A\&A}
\newcommand{\aj}{AJ}
\newcommand{\apj}{ApJ}
\newcommand{\apjl}{ApJL}
\newcommand{\araa}{ARA\&A}
\newcommand{\chjaa}{ChJAA}
\newcommand{\mnras}{MNRAS}
\newcommand{\nat}{Nature}
\newcommand{\pasj}{PASJ}

\title[On the relevance of chaos]{On the relevance of chaos for halo stars in the Solar
Neighbourhood}
\author[N. P. Maffione et al.]{N. P. Maffione$^{1,2}$\thanks{E-mail: nmaffione@fcaglp.unlp.edu.ar}, F. A. G\'omez$^{3}$, P. M. Cincotta$^{1,2}$, C. M. Giordano$^{1,2}$, \newauthor A. P. Cooper$^{4}$, B. W. O'Shea$^{5,6,7,8}$\\
$^{1}$Grupo de Caos en Sistemas Hamiltonianos Multidimensionales, FCAG-UNLP, La Plata, Buenos Aires, Argentina\\
$^{2}$Instituto de Astrof\'isica La Plata, UNLP-CONICET, La Plata, Buenos Aires, Argentina\\
$^{3}$Max-Planck-Institut f\"ur Astrophysik, Karl-Schwarzschild-Str. 1, D-85748, Garching, Germany\\
$^{4}$Institute for Computational Cosmology, Department of Physics, University of Durham, South Road, Durham, DH1 3LE, UK\\
$^{5}$Department of Physics and Astronomy, Michigan State University, East Lansing, MI 48824, USA\\
$^{6}$Institute for Cyber-Enabled Research, Michigan State University, East Lansing, MI 48824, USA\\
$^{7}$Lyman Briggs College, Michigan State University, East Lansing, MI 48825, USA\\
$^{8}$Joint Institute for Nuclear Astrophysics, Michigan State University, East Lansing, MI 48824, USA}

\begin{document}

\date{}

\pagerange{\pageref{firstpage}--\pageref{lastpage}} \pubyear{2002}

\maketitle

\label{firstpage}

\begin{abstract}
We show that diffusion due to chaotic mixing in the
Neighbourhood of the Sun may not be as relevant as previously suggested in
erasing phase space signatures of past Galactic accretion events. For this
purpose, we analyse Solar Neighbourhood--like volumes extracted from
cosmological simulations that naturally account for chaotic orbital
behaviour induced by the strongly triaxial and cuspy shape of the resulting
dark matter haloes, among other factors. In the approximation of an 
analytical static triaxial model, our results show that a large
fraction of stellar halo particles in such local volumes have chaos onset
times (i.e., the timescale at which stars commonly associated with chaotic orbits 
will exhibit their chaotic behaviour) significantly larger than a Hubble time. 
Furthermore, particles that do present a chaotic behaviour within a Hubble time do not 
exhibit significant diffusion in phase space.
\end{abstract}


\section{Introduction}
\label{sec:introduction}

In galactic dynamics, the term chaos refers to the exponential divergence of initially nearby orbits in phase space. In non--integrable systems, initially nearby stars in strong chaotic regions drift away from each other very quickly, thus losing memory of their common origin \citep[e.g.,][and references therein]{MV96,Co02,EVK08}. Understanding whether this physical process plays a major role in shaping the stellar phase space distribution of the Solar Neighbourhood is of key importance, as it is within this distribution that we hope to gain a significant insight into the formation history of the Galaxy \citep{HZ00,J08,GHBL10}.

The characterisation of the formation history of our own Galaxy is a very ambitious undertaking by modern astronomy \citep{FBH02}. Understanding how the Milky Way evolved to become the galaxy we currently inhabit would allow us not only to explore our origins, but also to understand galaxy formation in a more general context \citep{H08}. More precisely, it can allow us to test the current paradigm of galaxy formation and evolution. This theory predicts that the present--day population of galaxies grew in mass by merging with smaller companions. As their potential wells grew deeper, galaxies continued to accrete gas that cooled, collapsed into a disk, and gave rise to most of their stellar component \citep{WR78}. In addition to this in--situ population, every galaxy is predicted to have a minor fraction of its stellar content associated with merger events \citep{SZ78}. The tidal force that a satellite experiences as it orbits its host may be strong enough to disrupt it significantly. As a result of this interaction, initially spatially coherent and extended stellar streams are formed \citep[e.g.][]{IGI94,ILITQ01,bj05,Betal06,Betal07,Cetal10}. These streams are thus fossil signatures of this formation process and their identification is key to reconstructing the merger history of our Galaxy.

Stellar streams associated with the most ancient accretion events are expected to populate the inner Galactic regions, in particular the Solar Neighbourhood \citep{HWS03,Getal13}. Unfortunately, dynamical times in these regions are relatively short and streams tend to mix rapidly, losing their spatial coherence \citep{HW99,H08}. Even though galaxies are essentially collisionless \citep{BT87}, and thus streams become more clustered in velocity as they diffuse in configuration space, statistically significant phase space overdensities are needed to identify such streams in the Solar Neighbourhood. Clearly, the longer it takes for a stream to diffuse in configuration space, the larger the chances of identifying it in phase space. The rate at which a stellar stream spatially dissolves depends not only on the orbit of its progenitor satellite, but also on the properties of the host galactic potential \citep{HW99,H08,VWHS08}. In a galactic potential where a stream is on a regular orbit, its local density decreases in time as a power law, thus relatively slowly. Yet the current paradigm of galaxy formation predicts strongly triaxial and cuspy dark matter haloes \citep[see e.g.][]{JS02,Aetal06,VCetal11}. These two characteristics are known to be a significant source of chaos in a galactic potential \citep{S93,MF96,Siopis2000,VKS02,KS03,KVC05,MCW05}. On a chaotic orbit, the local stream density decreases at an exponential rate. Furthermore, chaotic orbits can significantly drift in the space of quantities that are otherwise approximately conserved, such as angular momentum \citep[e.g.][]{poveda92,schall97,Vetal13}, thus hindering its identification at the present day.

One of the main goals of ongoing and past astrometric, photometric and spectroscopic missions is to map the stellar phase space distribution in the Solar Neighbourhood. In addition to  {\it Gaia}  \citep{Petal01,Letal08}, which will measure positions and velocities of more than a thousand million stars, several other projects have provided and will continue to provide complementary information \citep[e.g.][]{zhao06,zwitter08,hermes,apogee,G12,desi,T14}. A meaningful interpretation of the degree of substructure found in the Solar Neighbourhood requires a deep understanding of the role that chaotic mixing plays in shaping the underlying substructure's phase space distribution. 

Since the identification and characterisation of chaotic orbits is a fundamental step towards this goal, efficient and accurate tools for this purpose are essential. A seminal contribution to the field of chaos
detection was made by Lyapunov \citep{L92} when he introduced the idea behind the so--called Lyapunov Characteristic Exponents (LCEs). LCEs are theoretical quantities that provide a measure of the rate of local exponential divergence of two initially nearby orbits in phase space. Thus, the LCEs are a very convenient way to distinguish between chaotic and regular motion and, particularly, to characterise chaos. Of particular importance is the largest LCE (lLCE), which is the LCE in the direction for which these two orbits diverge most rapidly. Theoretically, the characterisation of an orbit according to its lLCE is done based on its asymptotic behaviour at infinity. The Lyapunov Indicator (LI), on the other hand, refers to the finite--time version of the lLCE. A numerical value of the LI very close to zero indicates regular behaviour whereas any positive value indicates chaotic motion. The inverse of the LI provides a measure of the timescale for the manifestation of the exponential divergence. In practice, numerical finite--time techniques based on the concept of local exponential divergence like the LI \citep[see e.g.][]{BGGS80,S10} are commonly considered to be chaos indicators.

Nowadays, there are many chaos indicators in the literature that were developed based on the idea of the LCEs. Among the most used and tested indicators we find the Fast Lyapunov Indicator \citep[FLI;][]{FGL97,FL00,LF01,GLF02,LGF10}. The reliability shown by the FLI in previous works  \citep[e.g.][]{MGC11,MDCG11,DMCG12,MDCG13} makes this chaos indicator an ideal tool to characterise the role of chaos in Solar Neighbourhood--like volumes.

In this work we will take advantage of a variant of the FLI, the so--called Orthogonal Fast Lyapunov Indicator \citep[OFLI;][]{FLFF02} to evaluate the importance of chaos and chaotic mixing in shaping the phase space distribution of halo stars in the Solar Neighbourhood. 

The paper is organised as follows: we describe the models and the techniques in Section~\ref{sec:methodology} and include a short but comprehensive characterisation of the behaviours of the OFLI in Section~\ref{sec:meth-ind-bridge}. In Section~\ref{sec:relevance} we present our results on the actual impact of chaos in erasing local signatures of accretion events on the stellar halo phase space and revise the widespread assumption that chaos must inevitably lead to diffusion. We discuss and summarise our results in Section~\ref{sec:discussion}. Finally, in the Appendix we include a brief introduction to the concept of local exponential divergence, the definition of the LI and a more formal description of the phenomenon of stickiness and sticky orbits.  

\section{Methodology}
\label{sec:methodology}

In this Section we introduce the simulations and numerical tools used to characterise the chaotic nature of the stellar halo phase space distribution enclosed within Solar Neighbourhood--like volumes.

\subsection{Simulations}
\label{subsec:meth-sim}
  
We use five high--resolution, fully cosmological $N$--body simulations of the formation of Milky Way (MW)--like dark matter (DM) haloes carried out using the parallel Tree--PM code GADGET--3 \citep[an upgraded version of GADGET--2,][]{S05} by the {\it Aquarius Project} \citep[][]{Setal08a,Setal08b}. Each  halo was first identified within a large cosmological periodic box of 100 $h^{-1}$ Mpc a side \citep{BKSWJL09} and then re--simulated using a multi mass particle zoom--in technique. These DM--only simulations were performed using the following cosmological parameters: matter (dark and baryon) density, $\Omega_{m}=0.25$; dark energy density, $\Omega_{\Lambda}=0.75$; normalisation of the power matter spectrum, $\sigma_{8}=0.9$; scalar spectral index, $n_{s}=1$ and Hubble constant, $H_{0}=100~h$ km s$^{-1}$Mpc$^{-1} = 73$ km s$^{-1}$Mpc$^{-1}$, consistent with WMAP 1-- and 5--year constraints \citep{Spetal03,Ketal09}.
The DM haloes for the {\it Aquarius Project} were selected to have masses $\sim10^{12}$M$_{\odot}$, comparable to the MW, and to be relatively isolated at $z=0$. They were identified using a Friends--of--Friends \citep[][]{DEFW85} algorithm and self--bound subhaloes were identified with {\rm SUBFIND} \citep[][]{Setal01}. Each MW--like DM halo was re--simulated at a series of progressively higher resolutions. The experiments presented in this work are based on the simulations with the second highest resolution available, and their main properties are presented in Table~\ref{table:aquarius}. For a more detailed description of the simulations, we refer the reader to \citet{Setal08a,Setal08b}. 

\begin{table*}
\centering
\begin{minipage}{138mm}
\centering
\caption{Main properties of the five {\it Aquarius} haloes at $z=0$ from \citet{Setal08b}. The first column labels the simulation. From left to right, the columns give the virial radius of the DM halo, $r_{200}$; the virial mass, $M_{200}$; the number of particles within $r_{200}$, $N_{200}$; the particle mass, $m_{\rm p}$; the concentration parameter, $c_{\rm NFW}$; the intermediate to major, $b/a$ and the minor to major, $c/a$, principal axial ratios computed using DM particles located within 6 to 12 kpc; the total stellar halo mass, $M_{*}$ \citep[our stellar halo masses also includes the mass assigned to the bulge component in][]{Cetal10} and the half--light radius from \citet{Cetal10}, $r_{1/2}$.\newline Masses are in M$_{\odot}$; distances in kpc and velocities in km s$^{-1}$.}
\label{table:aquarius}
\begin{tabular}{@{}cccccccccc} \hline \hline Name & $r_{200}$ & $M_{200}$ & $N_{200}$ & $m_{\rm p}$ & $c_{{\rm NFW}}$ & $b/a$ & $c/a$ & $M_{*}$ & $r_{1/2}$ \\
& & [$10^{12}$] & [$10^{6}$] & [$10^{3}$] & & & & [$10^{8}$] & \\
\hline Aq--A2  & $245.88$ & $1.842$ & $135$ & $13.7$ & $16.19$ & $0.65$ & $0.53$ & $3.8$ & $20$\\
Aq--B2 & $187.7$ & $0.8194$ & $127$ & $6.4$ & $9.72$ & $0.46$ & $0.39$ & $5.6$ & $2.3$\\
Aq--C2 & $242.82$ & $1.774$ & $127$ & $14.0$ & $15.21$ & $0.55$ & $0.46$ & $3.9$ & $53$\\
Aq--D2 & $242.85$ & $1.774$ & $127$ & $14.0$ & $9.37$ & $0.67$ & $0.58$ & $11.1$ & $26$\\
Aq--E2 & $212.28$ & $1.185$ & $124$ & $9.6$ & $8.26$ & $0.67$ & $0.46$ & $18.5$ & $1.0$\\
\hline
\end{tabular}
\end{minipage}
\end{table*}

To model the formation and present day properties of their galactic stellar haloes, these simulations were post--processed with a semi--analytic model of galaxy formation \citep{Cetal10}. This semi--analytic model consists of a set of coupled differential equations describing the evolution of baryons and derives the mass accretion histories and phase space information of halo stars from the underlying $N$--body DM--only simulations. Processes such as star formation, AGN feedback, stellar winds and chemical enrichment are introduced in the model through differential equations that are controlled via a set of adjustable input parameters. The parameters were set to simultaneously match a range of observable quantities such as the galaxy luminosity functions in multiple wavebands \citep{BAetal05,BOetal06,Fetal08}.

The approach followed by \citet{Cetal10} uses the technique known as \emph{particle tagging}. The idea behind this technique is to assume that the most strongly bound DM particles in progenitor satellites can be used to trace the phase space distribution of their stars. At every time--step the $1\%$ most--bound DM particles were selected to trace any newly formed stellar population in each galaxy in the simulation. This fraction was set such that properties of the satellite population at $z=0$ are consistent with those observed for the MW and M31 satellites. As a result of this procedure, each tagged particle has a different final stellar mass associated to it. From now on we will refer to the tagged particles as stellar particles. The main properties of the resulting stellar haloes  are summarised in Table~\ref{table:aquarius}. Note that our stellar halo masses ($M_{*}$) also include the mass assigned to the bulge component in \citet{Cetal10}; see \citet[][hereinafter G13]{Getal13} and references therein for further details.

\citet{Cetal10} showed that particle tagging methods applied to simulations with sufficient resolution can be used to generate MW--like stellar haloes that reproduce various observables regarding the structure and characteristics of the Galactic stellar halo and its satellite population. Furthermore, as shown by G13, for most simulated stellar haloes, the measured velocity ellipsoids at 8 kpc are in good agreement with the estimate for the local Galactic stellar halo\footnote{The velocity ellipsoids at 8 kpc from the galactic centre were measured along the direction of the major axis of the DM--halo to increase the particle resolution.} \citep{CB00}. Nonetheless, it should be kept in mind that the dynamical evolution of the baryonic components of galaxies in these simulations is much simplified. As previously discussed in G13, this likely has an effect on, e.g. the efficiency of satellite mass--loss due to tidal stripping, or the satellite's internal structural changes due to adiabatic contraction, and possibly even on its final radial distribution \citep{Letal10,RSTH11,SM11,GSD13}. Recently, \citet{BAetal14} compared the stellar haloes formed in fully Smoothed Particle Hydrodynamics (SPH) simulations of galaxy formation  with DM--only simulations of the same initial conditions. They found that the resulting stellar haloes have different concentrations and internal structure due to the different kinematics that DM particles show with respect to their SPH counterparts. However, in the particle tagging scheme used by \citet{BAetal14} only one tagging operation is performed per satellite, at the time of its infall to the main halo, whereas \citet{Cetal10} tag stars continuously, as they are formed. The \citet{Cetal10} approach permits stars to diffuse in phase space within their parent satellite before its disruption, and thereby reproduces SPH results more closely than the tagging scheme tested by \citet{BAetal14} (Le Bret et al. submitted).

\subsection{The galactic potential}
\label{subsec:meth-pot}

The computation of chaos indicators (hereinafter CIs) to study the dynamics of our MW--like 
stellar haloes requires the integration of the equations of motion coupled with the first 
variational equations. The latter are used to track in time the evolution of the separation 
between initially nearby orbits in phase space (see Appendix~\ref{sec:CIsrevisited}). Due to the 
very high resolution of our $N$--body simulations, using frozen representations of the underlying 
galactic potential \citep[see for instance][]{Vetal13} becomes computationally expensive.
Methods to approximate the underlying potential based on series 
expansions can be extremely accurate 
\citep[e.g.][]{CB73,HO92,W99,KEV08,LJEF11,V13,MLHS14}. However, a very large number of expansion 
terms are needed in order to justify this approach. For instance, in \citet{LJEF11}, the authors 
used a halo expansion method to accurately fit the potential of the halo Aq--A2. Their analysis 
showed that a force accuracy of less than 1\% can be achieved using a series expansion that 
includes all moments up to $n,l = 20$. The resulting potential contains 8000 terms, rendering the derivation of 
the first variational equations unfeasible.  Instead, we chose to approximate each galactic potential 
with a suitable analytic model. Note that in this work we are dealing with pure DM simulations. 
Thus, the dynamics of the stellar particles are governed by the potential of the DM halo.

It is important to mention that, in this work, the $N$--body simulations are mainly used to extract the parameters of the underlying triaxial potentials and to sample the phase space distribution of Solar Neighbourhood--like volumes (see Section~\ref{subsec:meth-init}). In other words, our purpose is not to accurately characterise the impact of chaos in the {\it Aquarius} haloes themselves, but to obtain reasonable descriptions of these numerically simulated DM haloes to reflect in our results the expected stochasticity due to the morphological properties of this galactic component. 

As shown by \citet{Vetal12}, taking initial conditions from a self--consistent model and evolving them in a slightly different potential should increase the fraction of chaotic orbits in the sample. Thus, our results are likely to overestimate the role of chaotic mixing in the systems under study.

In \citet{NFW96,NFW97}, the authors introduced a spherical density profile that provides a reasonable fit to the mass distribution of DM haloes of galaxies in a very wide range of mass and redshift. However, it is now known that in a $\Lambda$--CDM cosmology DM haloes are not spherical as assumed by this potential. Instead, these are expected to be strongly triaxial and, furthermore, their shape is expected to vary as a function of galactocentric distance \citep[see e.g.][]{Aetal06,VCetal11}. Introducing into our analysis the triaxiality of the galactic potential is of key importance, as this is one of the main sources of chaotic behaviour together with the cuspy profile \citep{VKS02,KS03,KVC05,MCW05}. 

\citet{VWHS08} presented a triaxial extension of this profile that takes into account triaxiality and  radial variation in shape. The associated potential, $\Phi_{\rm TRI}$, can be described by

\begin{equation}
\label{eq:nfw_tri}
\Phi_{\rm TRI}=-\frac{A}{r_p}\ln \left( 1+\frac{r_p}{r_s}\right)\ ,
\end{equation}
where $A$ is a constant defined as 
\begin{displaymath}
A=\frac{G\,M_{200}}{\ln\left(1+c_{\rm NFW}\right)-c_{\rm NFW}/\left(1+c_{\rm NFW}\right)}\ ,
\end{displaymath}
with $G$ the gravitational constant, $M_{200}$ the virial mass of the DM halo and $c_{\rm NFW}$ the concentration parameter; $r_s=r_{200}/c_{\rm NFW}$ is a  scale radius with $r_{200}$ the virial radius. The triaxiality of this potential is introduced through $r_p$,
\begin{displaymath}
r_p=\frac{(r_s+r)r_e}{r_s+r_e},
\end{displaymath}
where $r$ is the usual galactocentric distance and $r_e$  an ellipsoidal radius defined as
\begin{displaymath}
r_e=\sqrt{\left(\frac{x}{a}\right)^2 + \left(\frac{y}{b}\right)^2 +
\left(\frac{z}{c}\right)^2}.
\end{displaymath}
The quantities $b/a$ and $c/a$ represent the intermediate to major and the minor to major principal axial ratios and are defined such that $a^2+b^2+c^2=3$. In all simulations, the ratios and directions of the principal axes were computed using DM particles located within 6 to 12 kpc. Their values are listed in Table~\ref{table:aquarius}. 

Note that, in this approximation used to represent the underlying potential of DM haloes, the potential shape changes from ellipsoidal to near spherical at the scale radius, $r_s$. Thus, for $r \ll r_s$, $r_p \backsimeq r_e$ and for $r \gg r_s$, $r_p \backsimeq r$ \citep{VWHS08}. We find that, to the $10\%$ level, these approximated analytic potentials can reproduce the true gravitational potentials within the relevant distance range, i.e., $r \lesssim 100$ kpc. This is in agreement with the recent results presented by \citet{Bonaca}, using the Via Lactea II simulation 
\citep{VLII}.

The potential $\Phi_{\mathrm{TRI}}$ admits, for $r_p<r_s$ the power series expansion

\begin{equation}
\Phi_{\mathrm{TRI}}=-\frac{A}{r_s}\sum_{n=1}^{\infty}\frac{{(-1)}^{n+1}}{n}
\left(\frac{r_p}{r_s}\right)^{n-1},
\label{expansion}
\end{equation}
so it is analytic everywhere, and the condition $r_p<r_s$ implies that $r,r_e<r_s$.

Under the above assumption, $r_p/r_s$ could be
approximated, up to $r_e^2/r_s^2$, by
\begin{displaymath}
\frac{r_p}{r_s}\approx\frac{r_e}{r_s} \left(1+\frac{r}{r_s}\right)\left(1-\frac{r_e}{r_s}\right).
\end{displaymath}
In spherical coordinates $(r,\vartheta,\varphi)$, introducing the parameters
\begin{displaymath}
\varepsilon_1=\frac{1}{8}\left(\frac{a^2}{b^2}-1\right),\quad \varepsilon_2=\frac{1}{4}\left(\frac{a^2}{c^2}-1\right),
\end{displaymath}
retaining terms up to $r_p/r_s$ in (\ref{expansion}) and neglecting a constant term, the potential takes the form

\begin{multline}
\Phi_{\mathrm{TRI}}(r,\vartheta, \varphi)\approx\Phi_0(r)+\Phi_1(r)\left\{\left(\varepsilon_2-{\varepsilon_1}\right)\cos2\vartheta-\right.\\
-{\varepsilon_1}\cos2\varphi+\frac{\varepsilon_1}{2}\cos2(\vartheta+\varphi)+\frac{\varepsilon_1}{2}\cos2(\vartheta-\varphi)\},
\label{phi_sphe}
\end{multline}

where

\begin{eqnarray*}
\Phi_0(r)&=&\frac{Ar}{2ar_s^2}\left(1+\frac{r}{r_s}\right)\left(1-\frac{r}{ar_s}\right)+(\varepsilon_1+\varepsilon_2)\Phi_1(r),\\\\
\Phi_1(r)&=&\frac{Ar^2}{2ar_s^2}\left(1+\frac{r}{r_s}\right)\left(\frac{1}{r}-\frac{2}{ar_s}\right).
\end{eqnarray*}

This approximation will be used in Section~\ref{subsec:diffusion} when discussing diffusion.

\subsection{Cosmological motivated initial conditions}
\label{subsec:meth-init}

To investigate the efficiency of chaotic mixing on halo stars in the vicinity of the Sun we first need to model their distribution in phase space. Rather than stochastically sampling the phase space distribution associated with the potential presented in Section~\ref{subsec:meth-pot}, Eq.~\eqref{eq:nfw_tri}, we select from each halo the stellar particles within spheres centred at 8 kpc from the corresponding galactic centre. Following \citet{GHBL10}, we choose for the spheres a radius of 2.5 kpc. This radius approximately corresponds to the distance within which the astrometric satellite {\it Gaia} will be able to measure with high accuracy positions and velocities of an extremely large number of stars. As the final configuration of the five host DM haloes is strongly triaxial, we have rotated each halo to its set of principal axes and placed the corresponding local spheres along the direction of the major axis. This allows a direct comparison between the different haloes. As shown by G13, varying azimuthally the location of our spheres results in local stellar densities that are, in general, an order of magnitude smaller than the observed value in the Solar Neighbourhood. Furthermore, the differences in number of resolved stellar streams within spheres  located at different azimuthal angles mainly reflect changes in the local stellar density. Thus, we do not expect our particular choice of location for these spheres to affect significantly our results concerning the dynamical nature of Solar Neighbourhood--like volumes. Finally, we will only consider stellar particles that originally were members of accreted satellite galaxies. Any stellar particle associated with in--situ star formation are disregarded.

\subsection{Chaos Indicator: The Orthogonal Fast Lyapunov Indicator}
\label{subsec:meth-ind-OFLI}

Now that the model and the volumes of interest have been introduced, the main goal of this section is to present briefly the preferred CI used in the analysis: the OFLI  \citep[][]{FLFF02}, a particular variant of the FLI\footnote{Even though we adopt the OFLI as our primary chaos indicator, this study is supported by similar results based on other CIs, such as the LI (a short introduction to the basic idea behind CIs and a definition of the most popular CI, the LI, are given in Appendix~\ref{sec:CIsrevisited}), the MEGNO \citep{CS00,CGS03,GKW05,CLD11}, the GALI \citep{SBA07,SBA08,MA11,MSA12} and the RLI \citep{SEE00,SESF04,SSEPD07,SESS04}. Thus, the orbital classification obtained with the approximate galactic potential described in Section~\ref{subsec:meth-pot} is very robust. However, for the sake of brevity the results based on these other CIs are not presented in this work.}.

Given an $N$--dimensional Hamiltonian $\mathcal{H}$. If we follow the time evolution of a unit
deviation vector $\mathbf{\hat{w}}(t)$ for a given solution of the equations of motion $\gamma(t)$,  initially chosen normal to the energy surface \citep[i.e. in the direction of  $\nabla\mathcal{H}$, see][]{B15}, take its orthogonal component to the flow at time $t$, $\hat{w}(t)^{\perp}\in\mathbb{R}$, and retain the largest value between an initial time $t_0$ and a stopping time $t_f$, we can define the OFLI as:

\begin{displaymath}
\mathrm{OFLI}^{\gamma}(t_f)=\sup_{t_0<t<t_f}\left[\hat{w}(t)^{\perp}\right]
\end{displaymath}
for the orbit $\gamma$. The $\mathrm{OFLI}^{\gamma}$ tends to
infinity as time increases for both non--periodic regular and chaotic orbits. The growth of
$\mathrm{OFLI}^{\gamma}$ is exponential with time if $\gamma$ is a chaotic orbit. The $\mathrm{OFLI}$ grows linearly with time for resonant and non--resonant regular orbits (on a logarithmic scale), with different rates, and oscillates around a constant value for periodic orbits \citep[for further details we refer the reader to][]{FGL97,FL00,LF01,FLFF02,GLF02,B15}. 

In what follows, we integrate the orbits and compute the CIs using the {\rm LP--VIcode} program. {\rm LP--VIcode} is a fully operational code which efficiently calculates a suite of 10 CIs in any number of dimensions \citep[see][]{CMD14}. The hardware we used for these experiments was an
Intel Core i5 with four cores, CPU at 2.67 GHz and 3 GB of RAM. The version of the \texttt{gcc gfortran} compiler was 4.8.2.

\section{Building up a basic understanding}
\label{sec:meth-ind-bridge}

Chaos indicators are used by dynamicists to identify and characterise the interplay between regular and chaotic components of diverse dynamical systems. For instance, CIs are a popular means of quantifying the impact of chaos on the dynamical evolution of self--consistent stellar systems, using non--evolving analytic models of the underlying potential \citep[][]{KV05,MA11,ZM12}. These CIs, such as the OFLI, provide a reliable and straightforward way to estimate the chaos onset times of orbital sets, i.e. the timescale at which stars commonly associated with chaotic orbits will effectively reveal their chaotic behaviour\footnote{Note that the chaos onset time is not the same as the commonly used Lyapunov time.}. In the following experiment we apply the OFLI to a regular orbit, a sticky orbit and a chaotic orbit in order to show what should be expected from the indicator in each case.

Sticky orbits are, for the purpose of this paper, chaotic orbits that behave regularly on timescales comparable to a Hubble time. In what follows, we classify an orbit as sticky if it exhibits chaotic behaviour only on timescales larger than 10 Gyr. For a more formal definition of sticky orbits, please see Appendix~\ref{sec:sticky}. 

\begin{figure*}
\begin{center}
\begin{tabular}{ccc}
\hspace{-5mm}\includegraphics[width=0.33\linewidth]{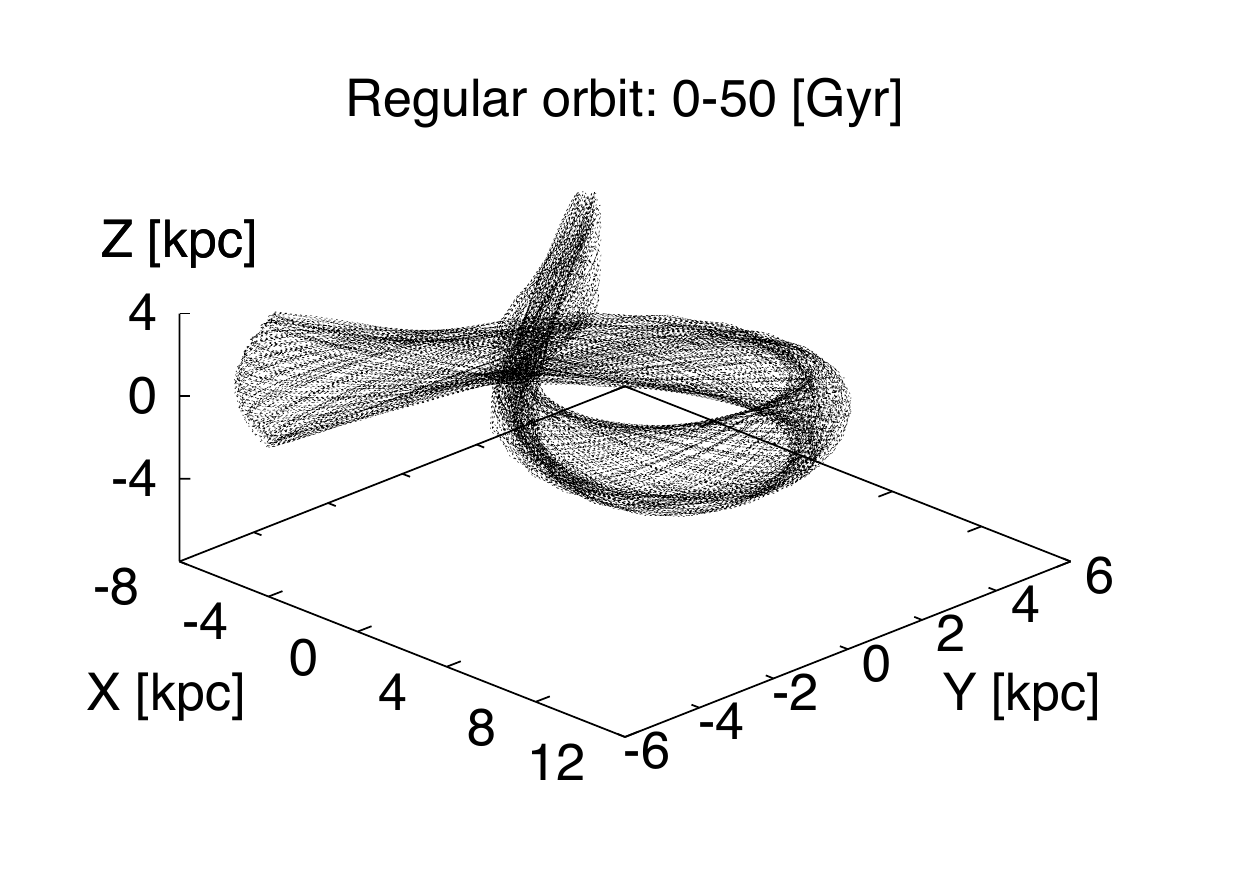}&
\includegraphics[width=0.33\linewidth]{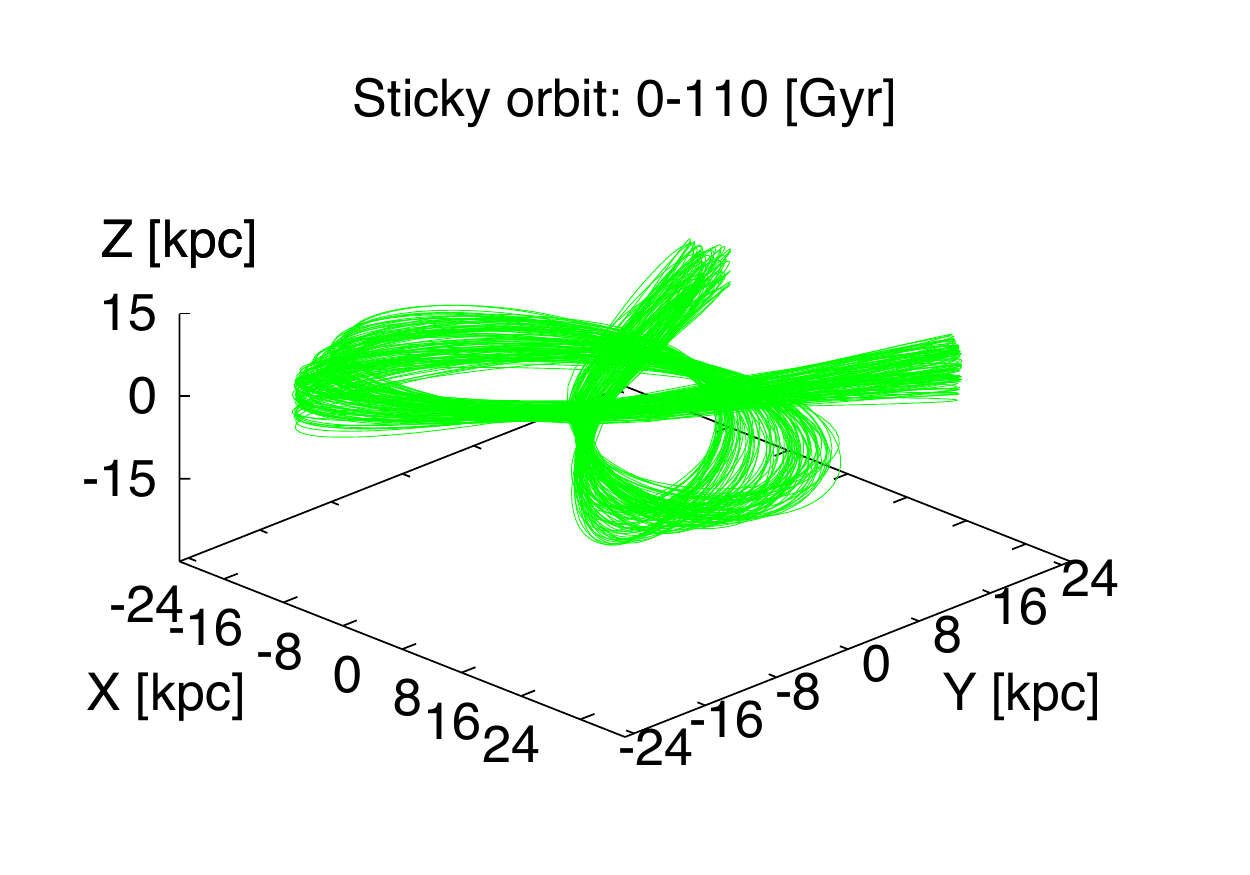}&
\includegraphics[width=0.33\linewidth]{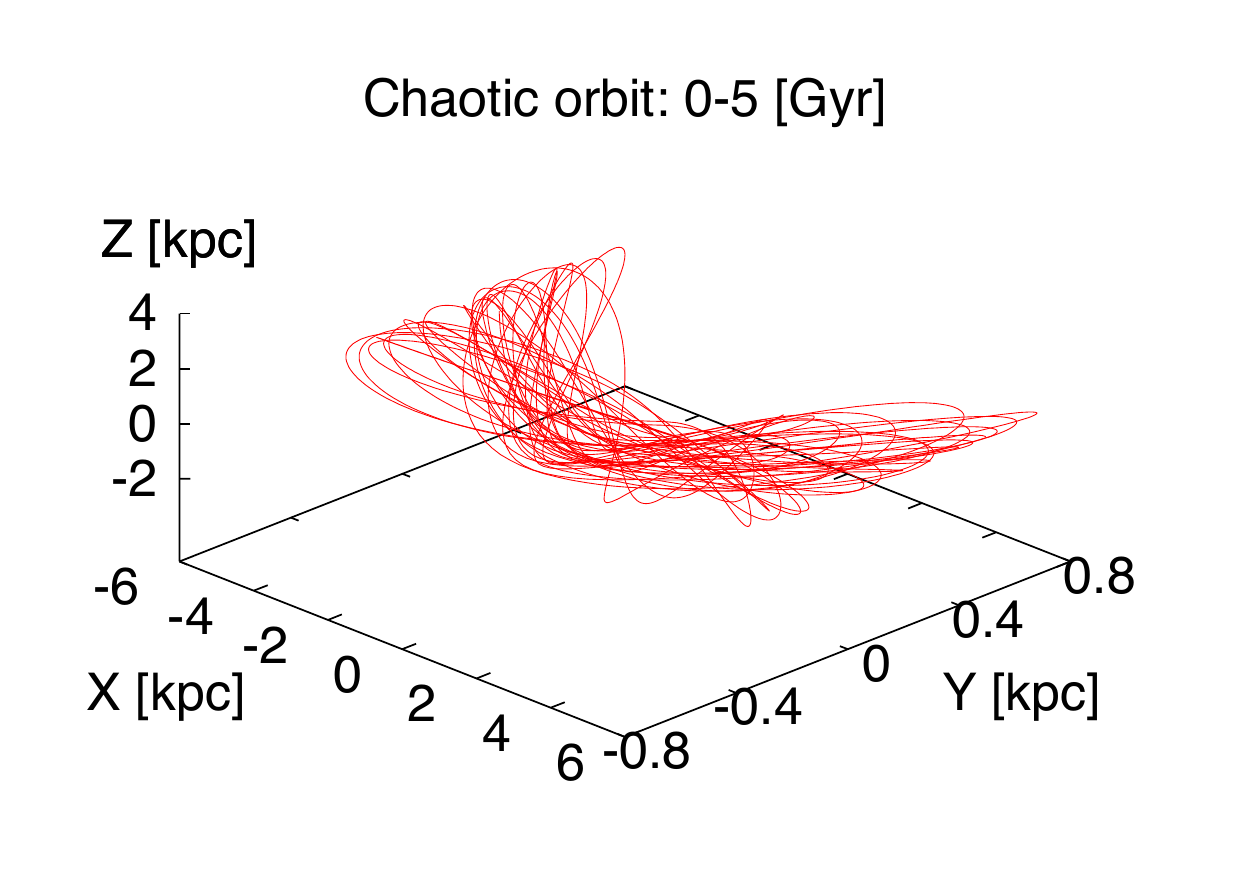}\\
\hspace{-5mm}\includegraphics[width=0.33\linewidth]{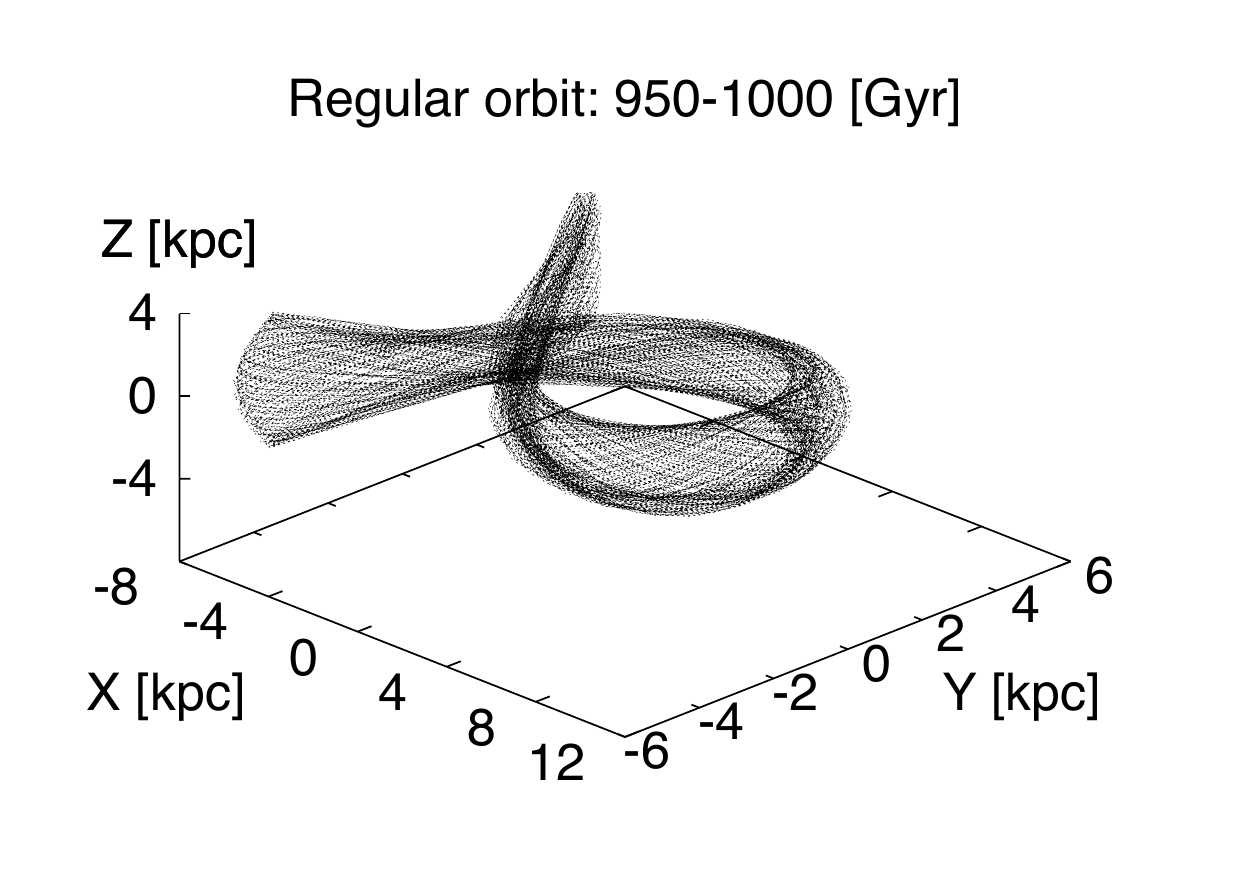}&
\includegraphics[width=0.33\linewidth]{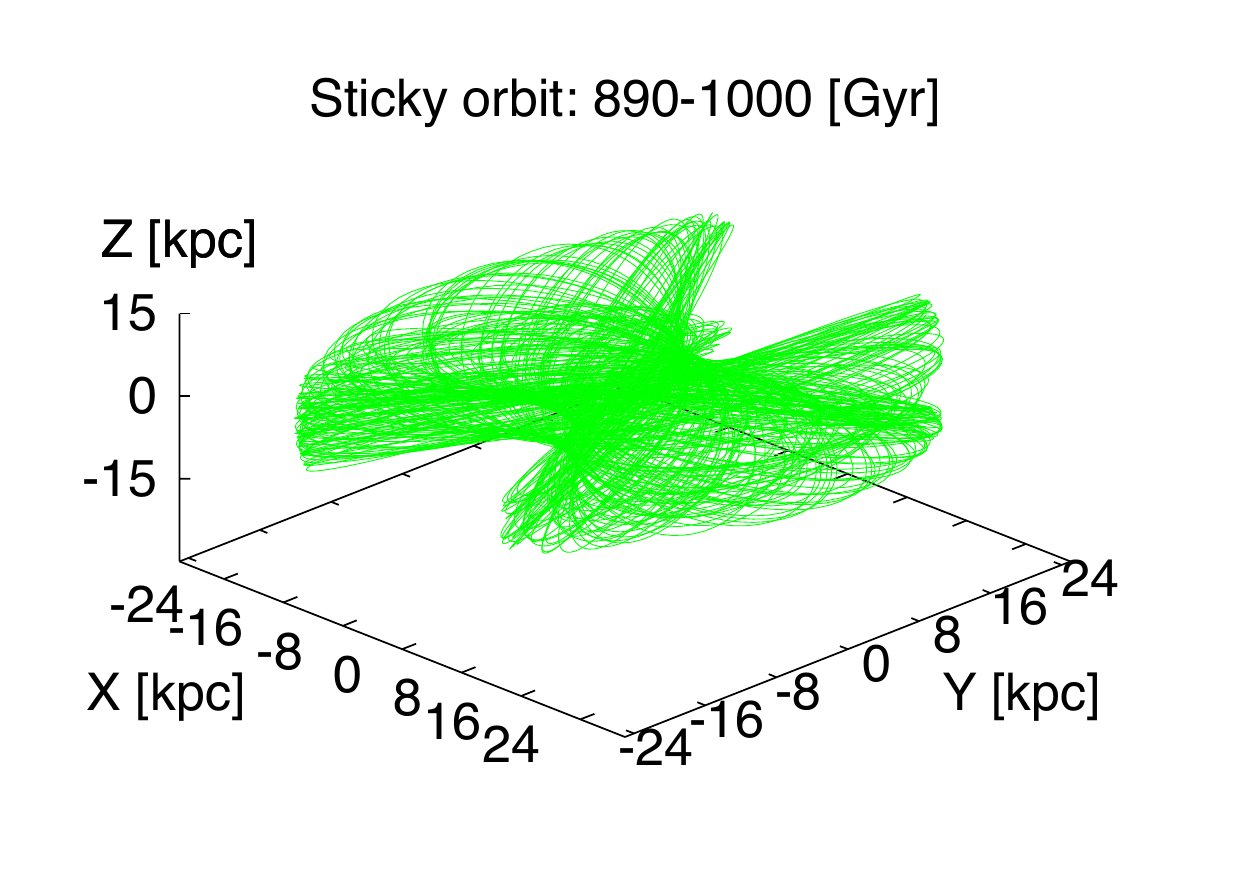}&
\includegraphics[width=0.33\linewidth]{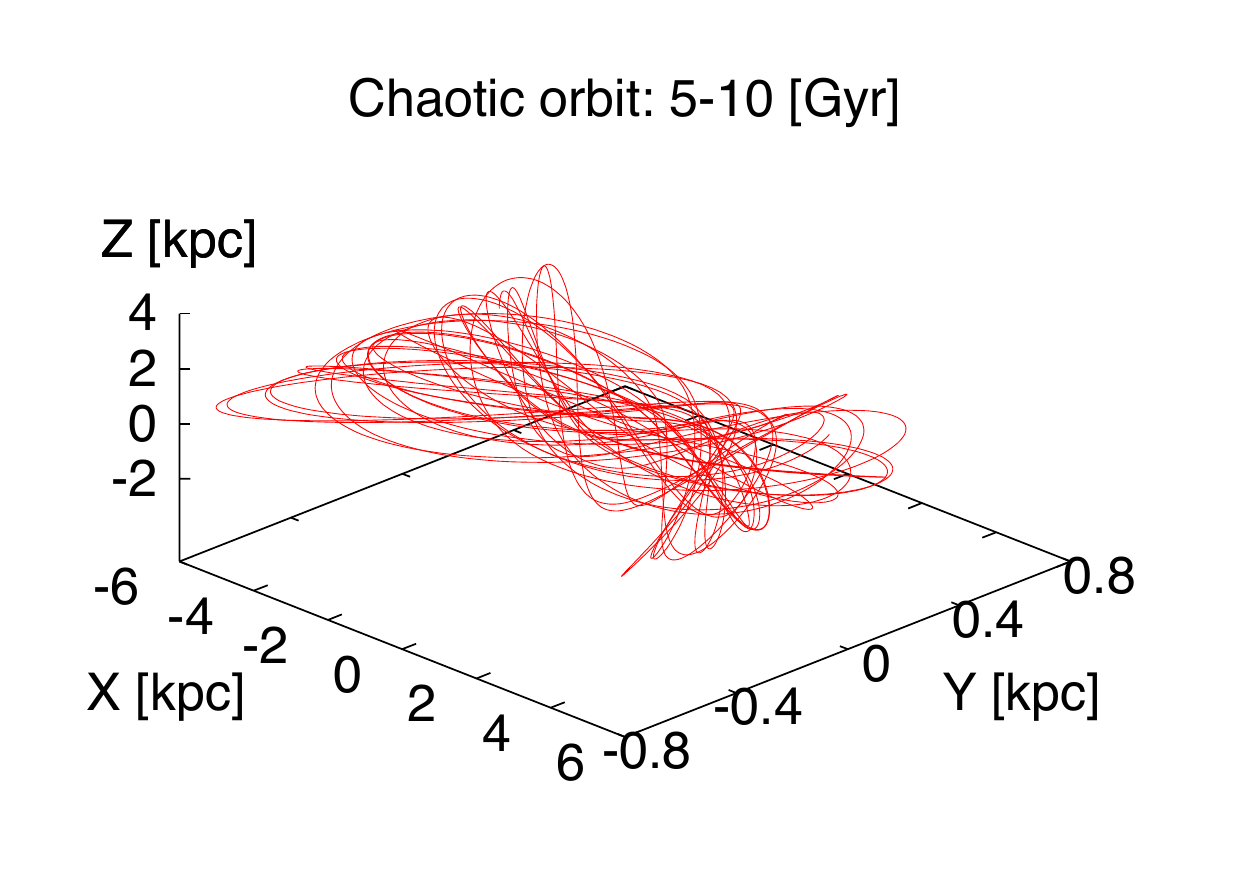}
\end{tabular}
\end{center}
\caption{Examples of regular orbits (left panels), sticky orbits (middle panels) and chaotic orbits (right panels) in the triaxial extension of the NFW model, for different time intervals. The fact that different volumes are occupied by the orbits in different time intervals (see the middle and right panels) is an indication of chaotic behaviour.}
\label{fig:000}
\end{figure*}

\begin{table*}
\centering
\begin{minipage}{138mm}
\centering
\caption{Initial conditions and binding energies for our examples of regular, sticky and chaotic orbits.\newline Distances are in kpc, velocities in km s$^{-1}$ and binding energies in km$^{2}$ s$^{-2}$.}
\label{table:init-examples}
\begin{tabular}{@{}cccccccc} \hline \hline Type of orbit & $x$ & $y$ & $z$ & $v_x$ & $v_y$ & $v_z$ & E\\
\hline Regular  & $8.219$ & $-0.652$ & $-2.203$ & $-5.795\times10^{-3}$ & $102.95$ & $-4.745$ & $-217691.706$\\
Sticky & $5.865$ & $0.263$ & $-0.346$ & $239.350$ & $333.547$ & $57.208$ & $-152469.329$\\
Chaotic & $5.731$ & $0.531$ & $0.443$ & $6.029$ & $-0.366$ & $23.691$ &  $-238489.404$\\
\hline
\end{tabular}
\end{minipage}
\end{table*}

In Fig.~\ref{fig:000} we present typical examples of regular orbits, sticky orbits and chaotic orbits in a triaxial extension of the NFW potential (Eq.~\eqref{eq:nfw_tri}) with the parameters of the DM halo Aq--A2 (Table~\ref{table:aquarius}). The (rounded) initial conditions for these orbits are presented in Table~\ref{table:init-examples}. We integrate simultaneously the equations of motion (Eqs.~\eqref{eq:3}) and the first variational equations (Eqs.~\eqref{eq:3.5}). The regular and sticky orbits are integrated for a rather long integration time of 1000 Gyr, while the chaotic orbit is integrated over only 10 Gyr, as this timescale is enough to characterise its behaviour. 
The time--step used for the regular and sticky orbits is $1$ Myr, and $0.01$ Myr for the chaotic orbit. The (rounded) binding energies (E) of the three orbits are included in Table~\ref{table:init-examples}. The integrator conserves energy to an accuracy of one part in $10^{-12}$ or less for all the experiments throughout the paper.

In Fig.~\ref{fig:000} we show only the first and last intervals of 50 and 110 Gyr for the regular and the sticky orbits, respectively, to illustrate the different behaviours that characterise these types of orbit in configuration space. The behaviour of the chaotic orbit is presented for two consecutive 5 Gyr intervals. 

From Fig.~\ref{fig:000}, it is clear that the regular orbit has a very similar shape in the two different time intervals, even though these are separated by 900 Gyr. Much more significant differences are apparent in the trajectories of the sticky orbit sampled at widely separated time intervals (0 to 110 Gyr for the middle top panel and 890 to 1000 Gyr for the middle bottom panel). This illustrates the chaotic nature of the sticky orbit. Nevertheless, it takes the orbit more than 8 Hubble times (roughly 110 Gyr) to exhibit its true nature. In the rightmost panels we show a chaotic orbit for two consecutive 5 Gyr intervals. Even in these much shorter intervals, the evolution in the trajectory is evident.

\begin{figure}
\begin{center}
\begin{tabular}{c}
\hspace{-5mm}\includegraphics[width=1\linewidth]{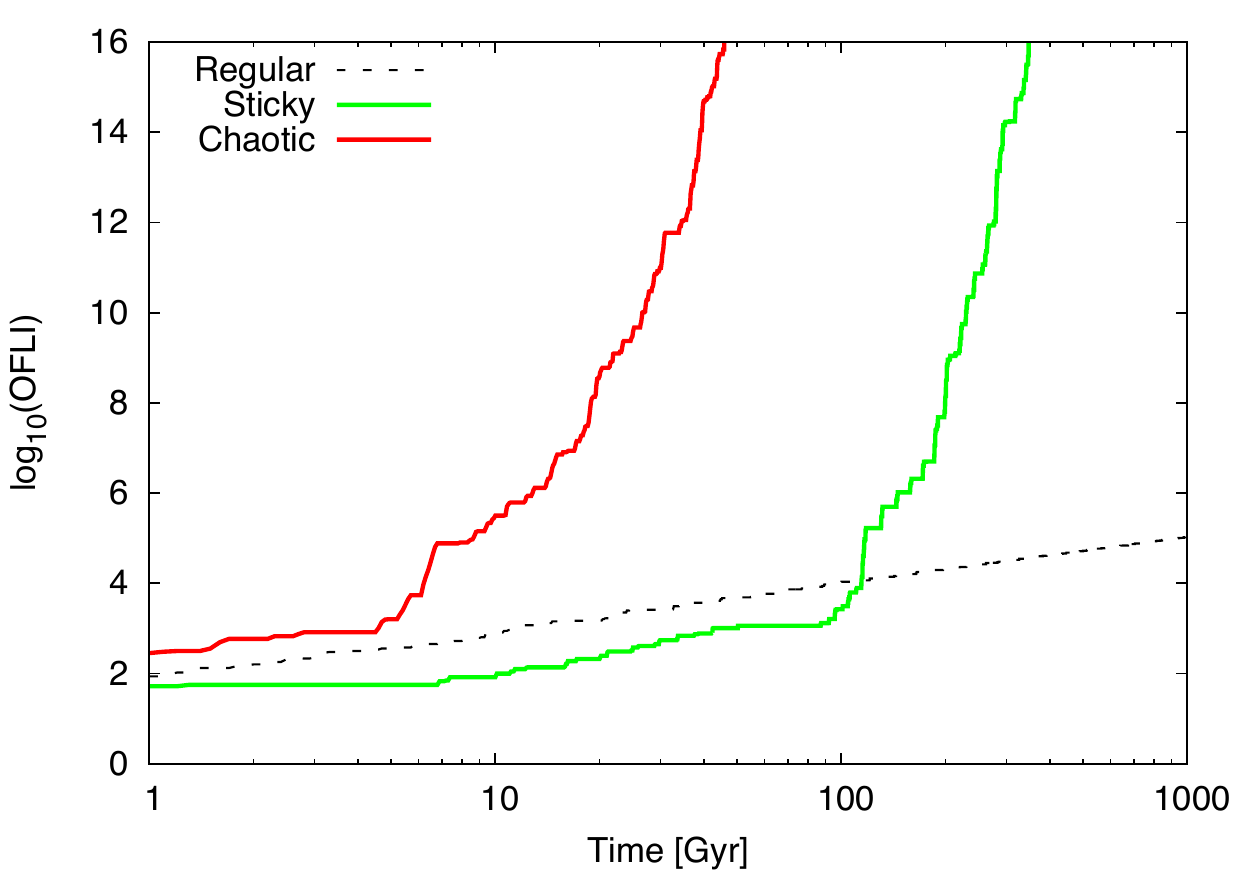}
\end{tabular}
\end{center}
\caption{Time evolution of the OFLI for the three orbits introduced in Fig.~\ref{fig:000}. Notice the logarithmic scale on both axes. The exponential growth of the indicator for chaotic motion is clearly observed for the chaotic and sticky orbits.}
\label{fig:00}
\end{figure}

In Fig.~\ref{fig:00} we present the characteristic behaviour of the OFLI for the three types of orbit showed in Fig.~\ref{fig:000}. In this case, all three orbits have been integrated using a time--step of 1 Myr. Notice the linear evolution of the indicator for regular orbits and the exponential growth corresponding to sticky and chaotic orbits. As expected from our previous discussion, for the sticky orbit, it takes the OFLI $\sim 100$ Gyr to start growing exponentially, time at which the chaotic behaviour of this particle is revealed. Instead, for the chaotic orbit, it only takes the OFLI a few Gyr to start showing an exponential growth. 

\section{The actual relevance of chaos: Solar Neighbourhood--like volumes}
\label{sec:relevance}

In order to characterise the impact of chaotic mixing on the phase space distribution of the Solar Neighbourhood, in this section we examine two central points: {\it i)} the distribution of chaos onset times for stellar particles  within Solar Neighbourhood--like volumes and {\it ii)} the rate of diffusion due to chaotic mixing, a mechanism that can lead to large variations of the integrals of motion. Our goal is to explore whether chaotic mixing can be strong enough to erase signatures of merger events in the neighbourhood of the Sun.

To tackle {\it (i)} we select particles from the five {\it Aquarius} stellar haloes in Solar Neighbourhood--like volumes (see Section~\ref{subsec:meth-init}). We then quantify the fraction of particles on regular, sticky and chaotic orbits. We do this by means of the OFLI, which allows us to estimate efficiently the distribution of chaos onset times and so identify the characteristic timescales over which chaotic mixing becomes relevant. To address {\it (ii)}, we measure the diffusion of pseudo--integrals of motion for large ensembles of test particles that are initially nearby in phase space. 

\subsection{The importance of timescales}

First we look for a relationship between the expected OFLI behaviours of different
orbit types (described above) and more readily interpreted measures of the local (stream) density. After this, we estimate chaos onset times to characterise the actual amount of chaos manifested on physically meaningful timescales. 

\subsubsection{The time evolution of the OFLI for initially nearby particles in phase space}
\label{subsubsec:bridge-connect}

In the previous section we have shown that our triaxial DM haloes admit a wide range of different orbital behaviours. In particular, we have shown an example of a very chaotic orbit with a chaos onset time shorter than 5 Gyr. These orbits can potentially play an important role in shaping the present--day phase space distribution of our Solar Neighbourhood. As shown by \citet{VWHS08}, the local  density in the neighbourhood of a particle moving on a chaotic orbit decreases exponentially with time. As a result, stellar streams moving on such orbits will experience a rapid phase space mixing process which can erase signatures of their accretion history. In contrast, the local density of a particle moving on a regular orbit decreases as a power law function of time, with exponent less than or equal to $3$ \citep[a significantly lower rate,][]{HW99,VWHS08,GHBL10}. Hence, the probability of finding streams on regular orbits in the Solar Neighbourhood may be somewhat higher.

With these ideas in mind, G13 followed the time evolution of the local density of accreted stellar particles that, at $z=0$, were located within different Solar Neighbourhood--like volumes. The goal was to explore whether particles that appeared to be smoothly distributed in phase space, and thus not associated with any resolved stellar stream, were on chaotic orbits. If not, the lack of clustering in phase space for these particles could be due to the limited numerical resolution of the simulations. G13 fit a power law function to the time evolution of the local density around each star particle and determined the rate at which this local density decreases with time. They assumed that a power law fit to the local density of stream particles on a chaotic orbit should yield an exponent greater than $3$. Unfortunately, due to the finite resolution of their simulations, the local volumes used to track the time evolution of density in G13 were rather large -- namely, spheres of radius equal to half of the apocentre of the particle's orbit. Such large volumes are not problematic if the goal is to detect resolved stellar streams. However, it is not clear that such fits reflect the true dynamical nature of the underlying local stream densities.

\begin{figure*}
\begin{center}
\begin{tabular}{ccc}
\hspace{-5mm}\includegraphics[width=0.33\linewidth]{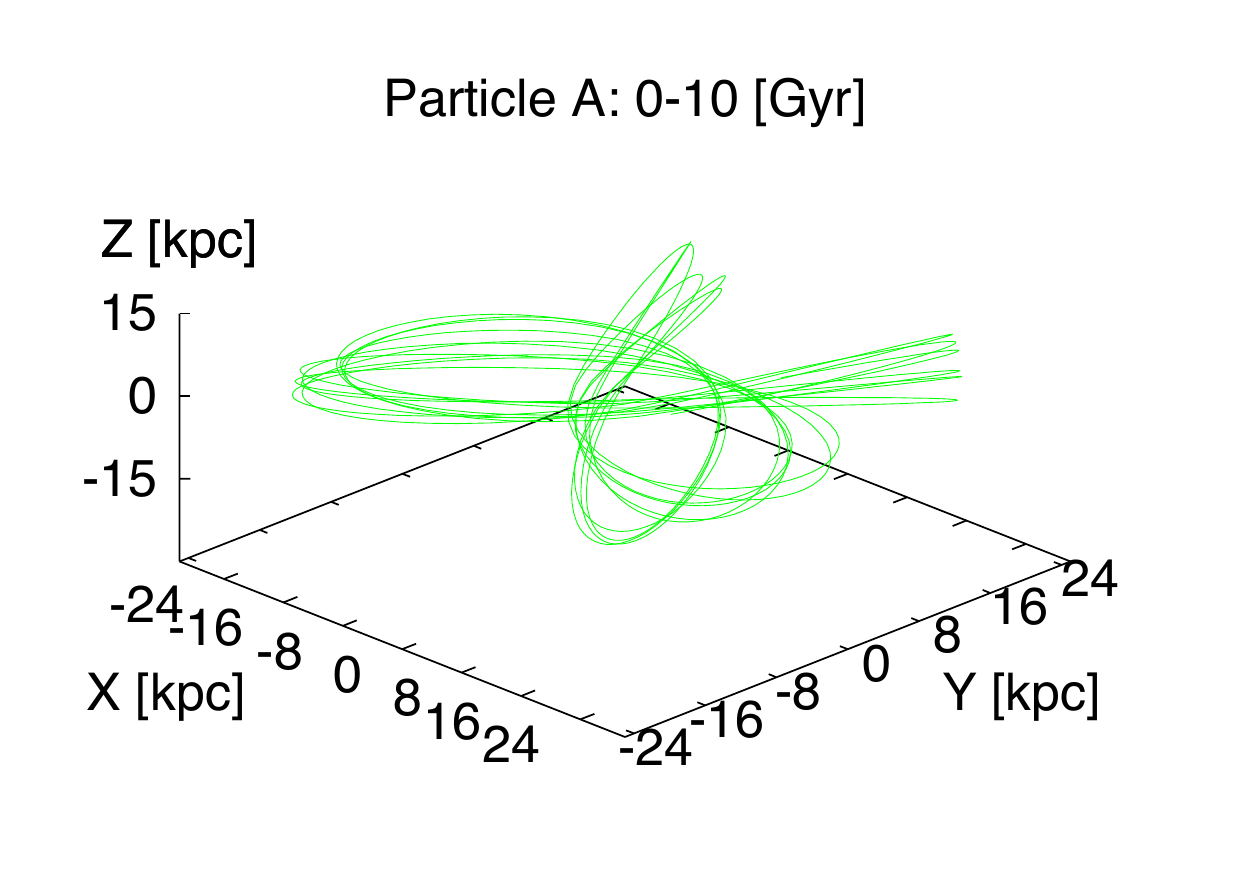}&
\includegraphics[width=0.33\linewidth]{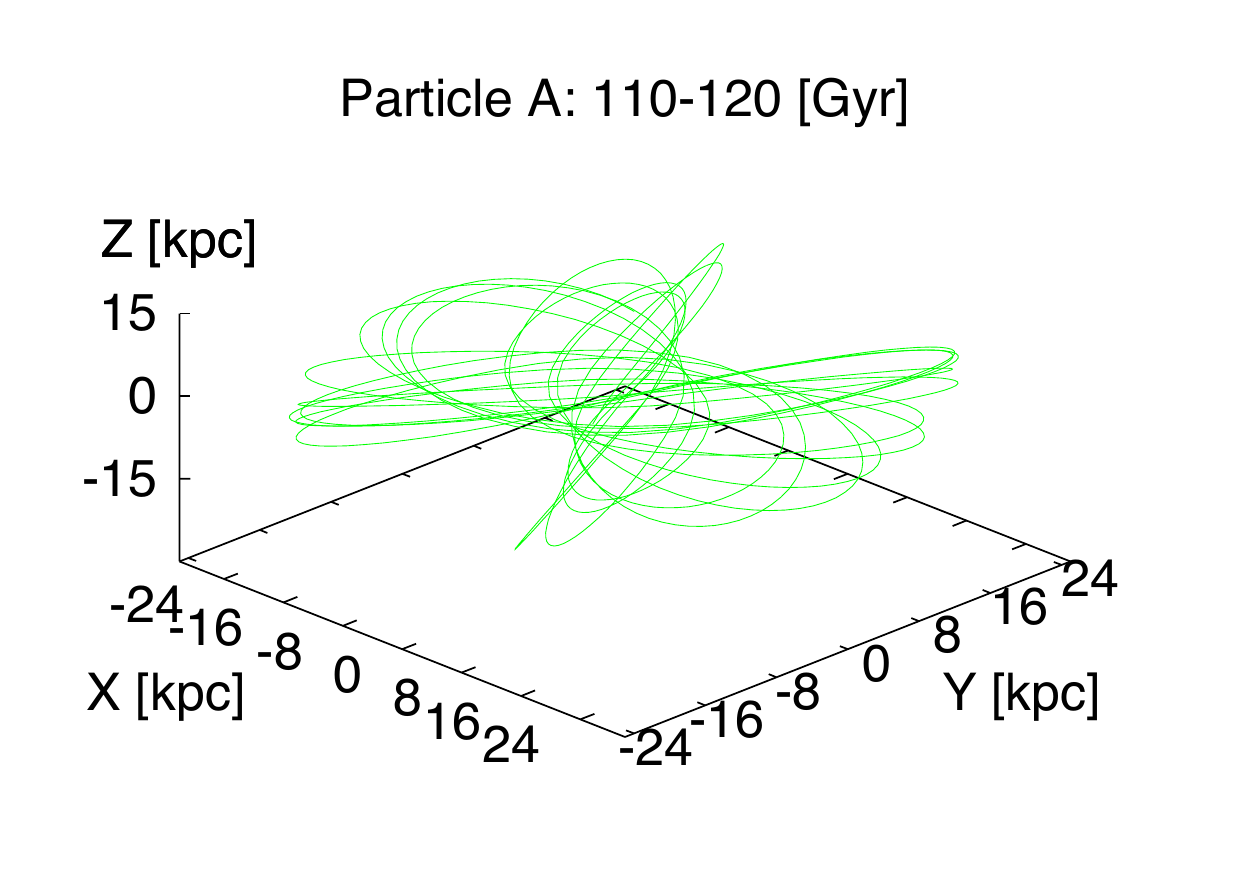}&
\includegraphics[width=0.33\linewidth]{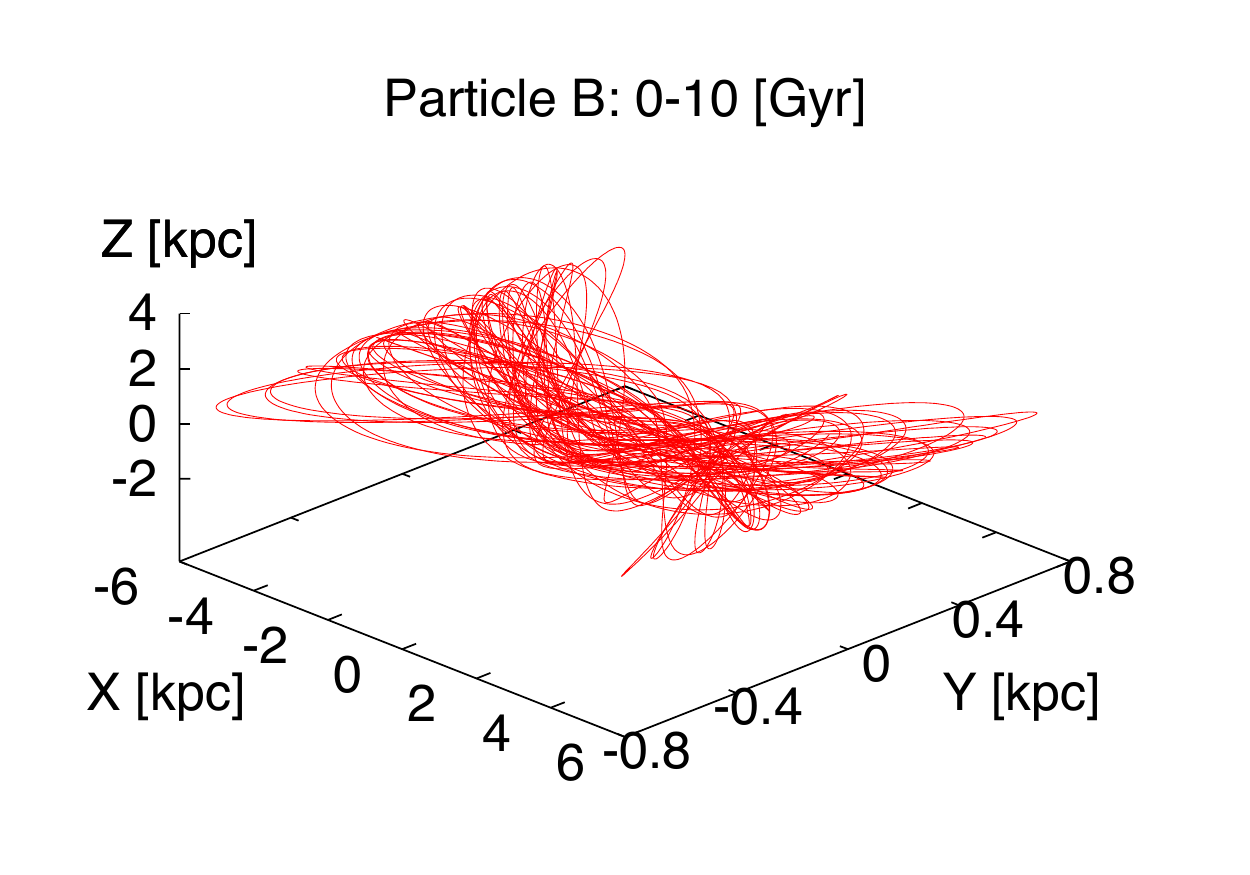}
\end{tabular}
\end{center}
\caption{Trajectories for both guiding particles (A and B) over different time intervals. The leftmost panel shows the trajectory of particle A from 0 to 10 Gyr and the central panel its trajectory from 110 to 120 Gyr. The rightmost panel shows the trajectory of particle B from 0 to 10 Gyr. The orbit associated with particle A has a chaos onset time larger than 110 Gyr. Thus, its shape only starts to change after that time. In case of particle B, the chaos onset time is shorter than 10 Gyr.}
\label{fig:0a*}
\end{figure*}

In the following experiments we explore the connection between the OFLI and the time evolution of the local density of a stream. We also explore the effects of the size of the local volume on the characterisation of the dynamical nature of a stream through the time evolution of its local density.

Figure~\ref{fig:0a*} shows the trajectories of two specific stellar particles in the Aq--A2 DM halo. These particles were chosen as clear examples of sticky and chaotic orbits, and at $z=0$ are found in a sphere of radius 2.5 kpc located at 8 kpc from the centre of the halo at $z=0$. We refer to these stellar particles as guiding particles A and B; their estimated chaos onset times are $\sim 113$ and 5 Gyr, respectively. 
To relate the OFLI to the time evolution of local density of streams, we distribute an ensemble of 1000 massless test particles in a small phase space volume around both guiding particles. These volumes are defined by multivariate Gaussians in phase space with initial dispersions $\sigma_{\rm x} = 0.2$ pc and $\sigma_{\rm v} = 1$ km/s. Test particles are initially distributed such that their maximum separation with respect to the guiding particle is less than or equal to $\sqrt{3}\sigma_{\rm x}$ and $\sqrt{3}\sigma_{\rm v}$ respectively. As we show below, such small phase space volumes are necessary to characterise accurately the dynamical behaviour of the local density around our particles A and B.

The ensembles of test particles and their corresponding guiding particles are then integrated in our triaxial NFW model (Aq--A2 DM halo parameters as Table~\ref{table:aquarius}) for 10 Gyr, with a timestep of 1 Myr. During the integration, the OFLI is computed for each particle (test and guiding). 

\begin{figure*}
\begin{center}
\begin{tabular}{ccc}
\hspace{-5mm}\includegraphics[width=0.33\linewidth]{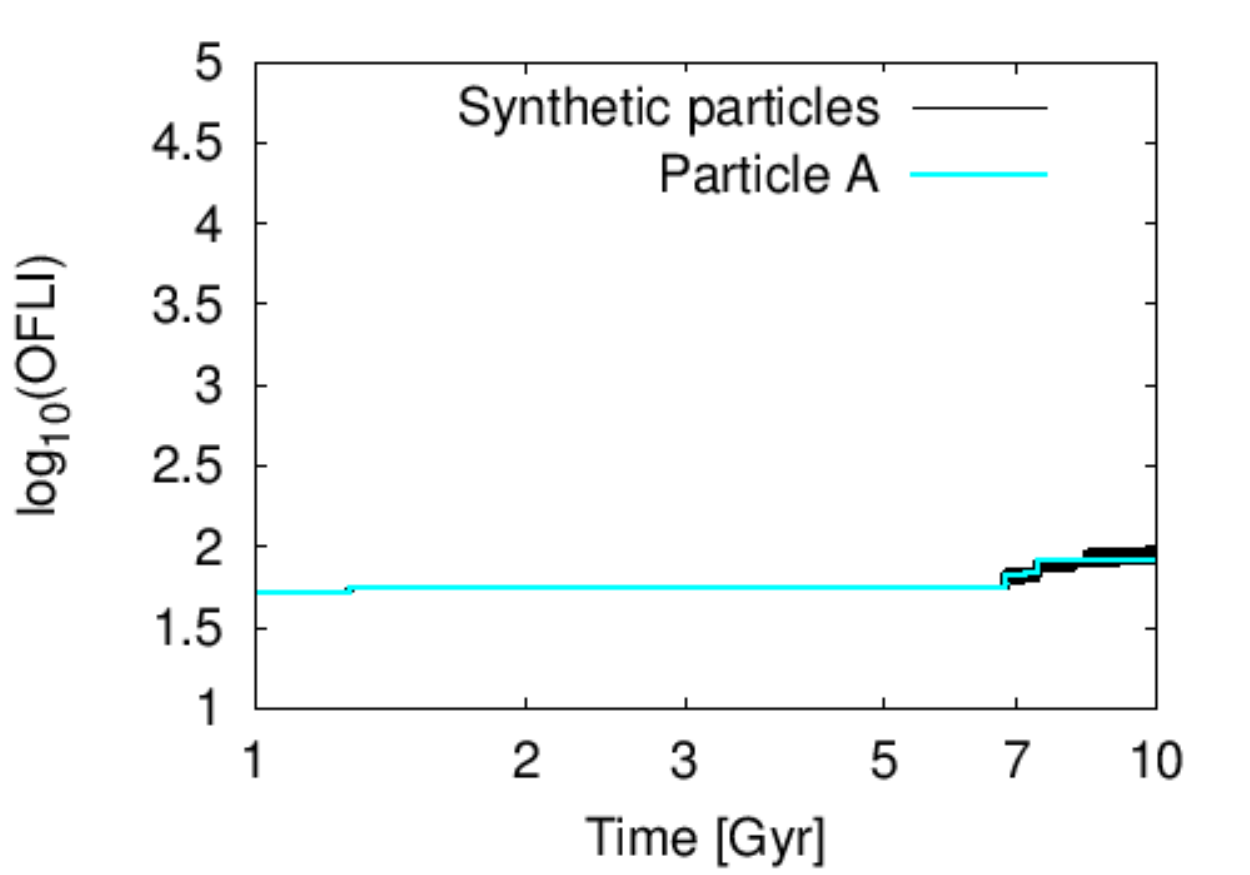}&
\includegraphics[width=0.33\linewidth]{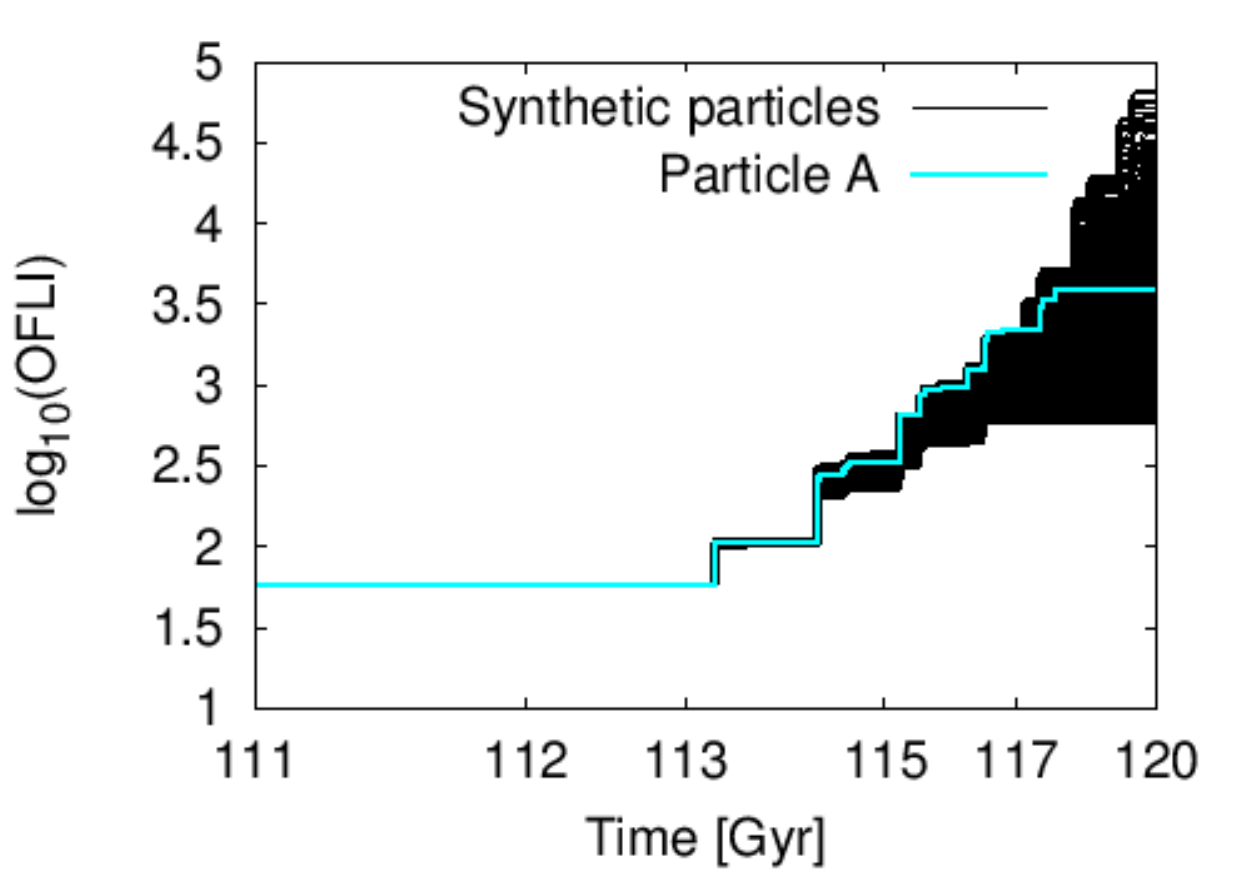}&
\includegraphics[width=0.33\linewidth]{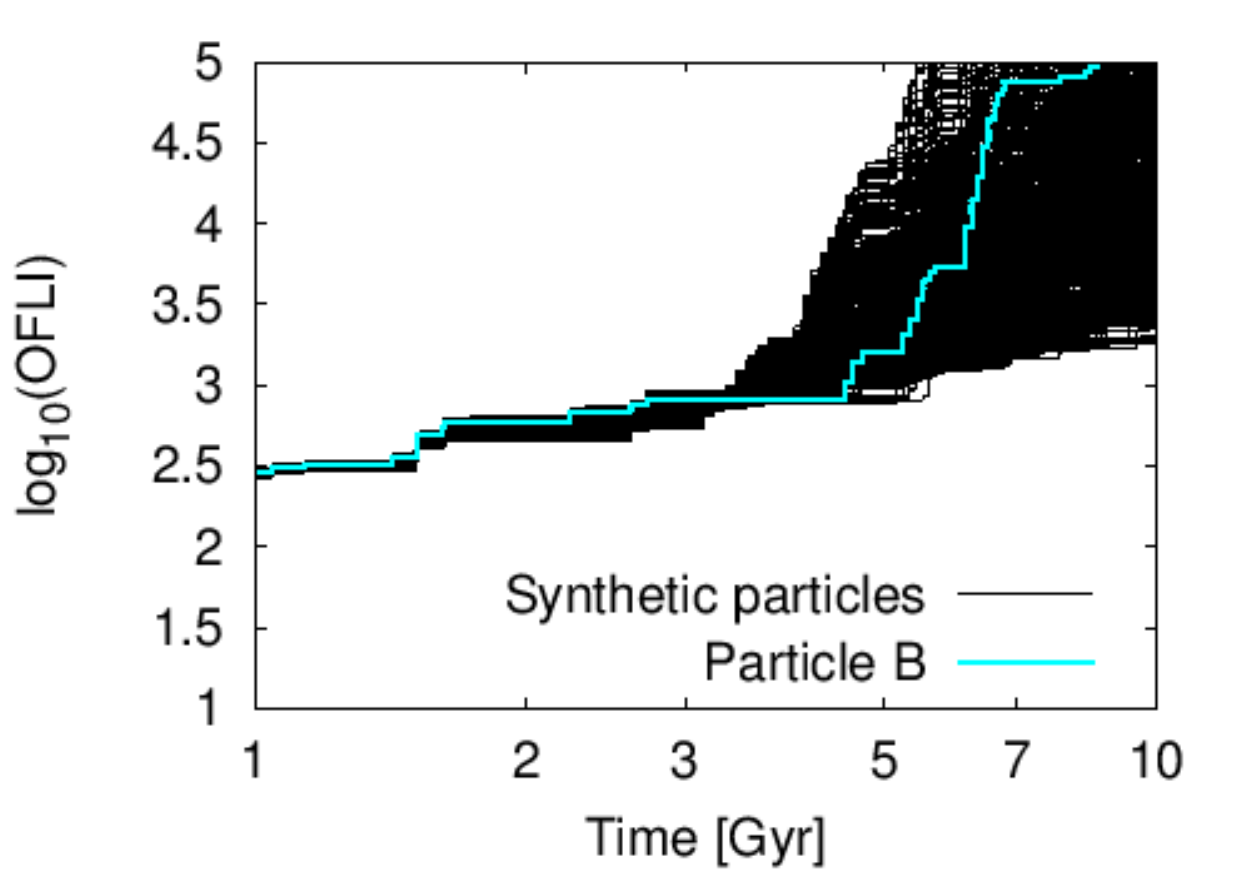}
\end{tabular}
\end{center}
\caption{Time evolution of the OFLI for both guiding particles (A and B) and their corresponding test particle ensembles (eA and eB). Guiding particles are shown in cyan and test particles in black. Left panel: OFLI for particle A and ensemble eA in the interval 0 to 10 Gyr. Middle panel:  OFLI for particle A and ensemble eA from 110 to 120 Gyr. Right panel: OFLI for particle B and ensemble eB in the interval 0 to 10 Gyr. Notice the logarithmic scale. The left and middle panels show the typical behaviour of the phase space surrounding a sticky orbit: regular behaviour followed by exponential growth of the OFLI at the chaos onset time. The last panel shows a much more rapid onset of exponential growth due to the high degree of chaos in the phase space neighbourhood of particle B.}
\label{fig:1}
\end{figure*}

The left and middle panels of Fig.~\ref{fig:1} show the time evolution of the OFLI for the guiding particle A and its associated ensemble of test particles, eA. Guiding particles are depicted in cyan and test particles in black. The time intervals shown in these panels are the same as those in Fig.~\ref{fig:0a*}, i.e. 0 to 10 Gyr (left panel) and 110 to 120 Gyr (middle panel). Note that separate ensembles of test particles are sampled from the Gaussian kernel defined above at the start of each interval shown.

\begin{figure*}
\begin{center}
\begin{tabular}{ccc}
\hspace{-5mm}\includegraphics[width=0.33\linewidth]{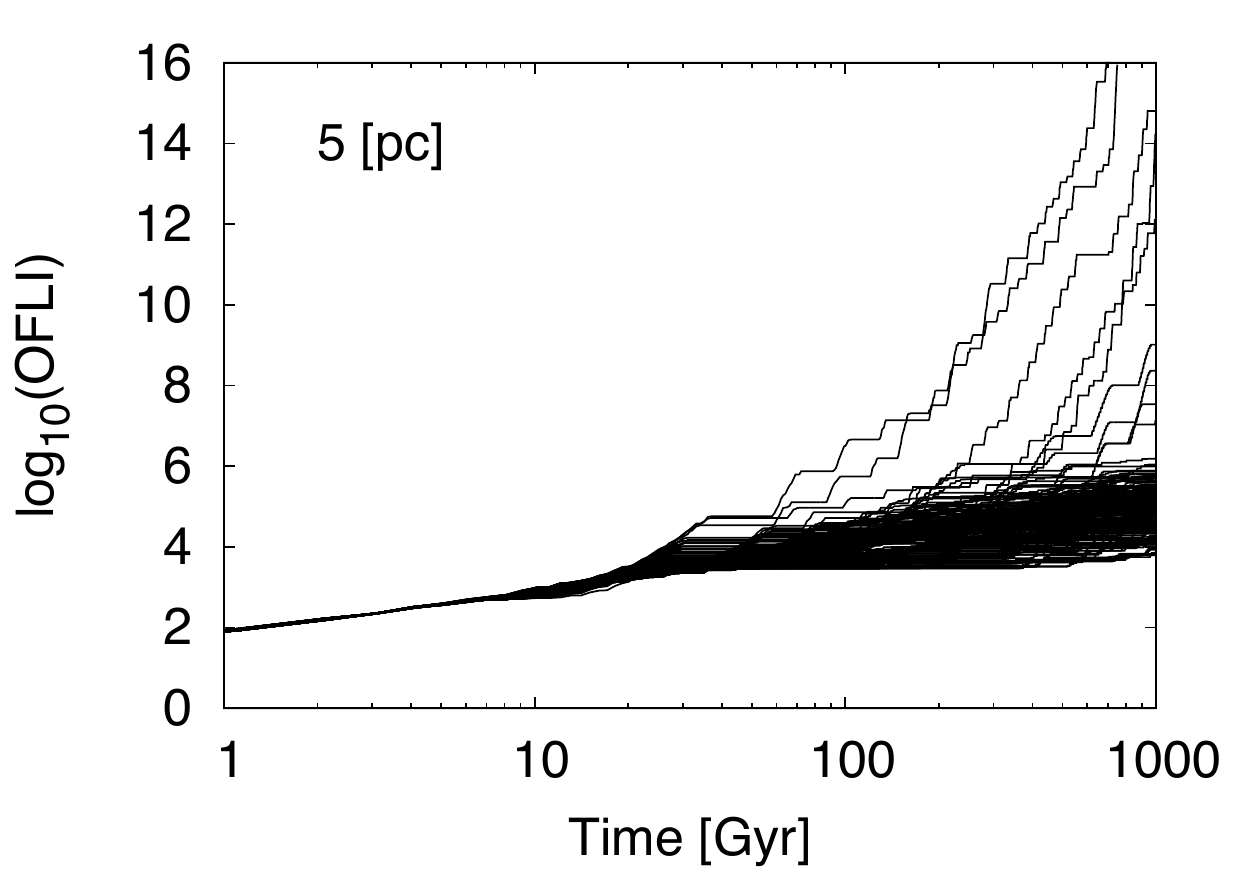}&
\includegraphics[width=0.33\linewidth]{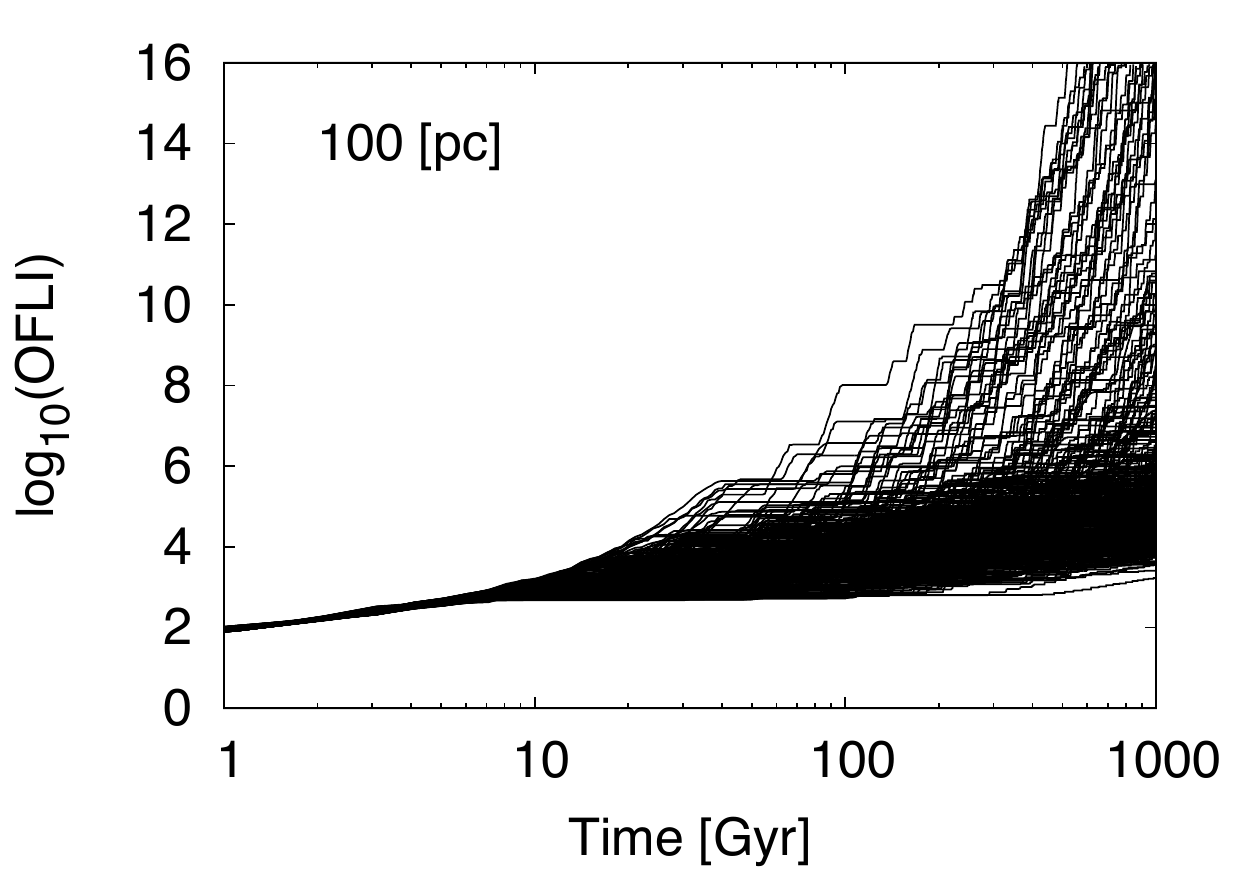}&
\includegraphics[width=0.33\linewidth]{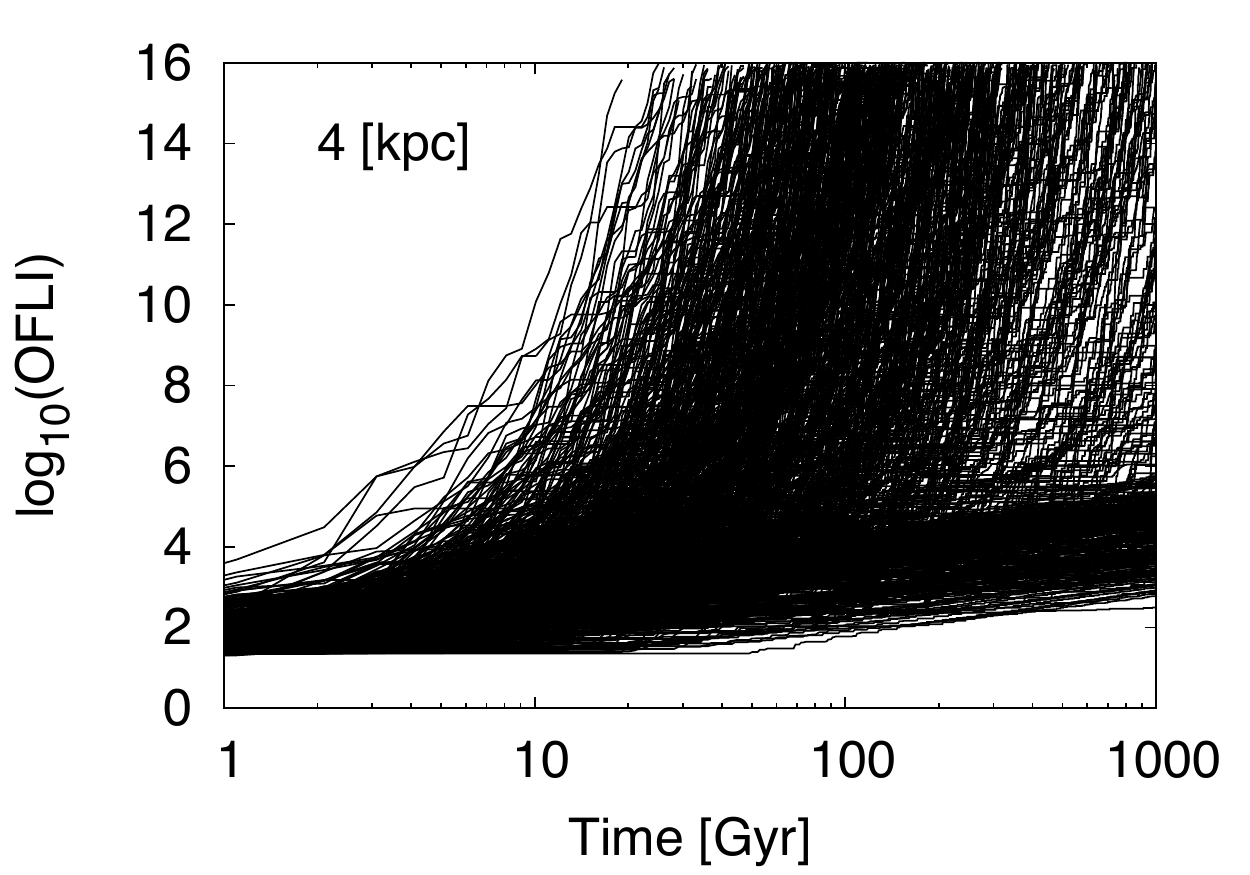}
\end{tabular}
\end{center}
\caption{Time evolution of the OFLI for 1000 test particle initial conditions sampled around a guiding particle on a regular orbit. The radius of the initial sphere is indicated. Notice the logarithmic scale. The bigger the sphere around the guiding particle, the more diverse the behaviours of the neighbouring particles.}
\label{fig:nosabe}
\end{figure*}

On the left panel of Fig.~\ref{fig:1}, we show that the OFLI for particle A  increases linearly with time over the first 10 Gyr of evolution, indicative of a regular orbit as described in Section~\ref{sec:meth-ind-bridge}. The indicator shows similar behaviour for all the test particles eA over the same interval (black solid curves). This indicates that the initial distribution of test particles accurately samples the phase space volume of the guiding particle.

The middle panel of Fig.~\ref{fig:1} shows results for another time interval, from 110 to 120 Gyr, for the same guiding particle A. The exponential growth of the OFLI in this interval (starting at $\sim113$~Gyr) implies a chaotic orbit. As before, we find similar behaviour of the OFLI for all the test particles in the corresponding ensemble, including the onset time of exponential growth. Clearly, particle A moves on a sticky orbit.
 
The right panel of Figure~\ref{fig:1} shows the time evolution of the OFLI for another guiding particle, B, with a chaotic orbit. Note that the chaotic nature of this particle's orbit becomes apparent on a much shorter timescale, $\sim 5$ Gyr. As in the second time interval shown for particle A, the vast majority of test particles in ensemble eB show rapid growth of the OFLI, reflecting the behaviour of their guiding particle. However, we find a much larger spread in the rate of the exponential growth in the case of particle B than in the case of the sticky orbit of particle A.

We have demonstrated that test particles  distributed within initially small phase space volumes around a guiding particle show very similar behaviour of the OFLI. This indicates that the time evolution of an ensemble of such particles can be used as an indirect indicator of the nature of any guiding particle. As previously described, this idea was exploited by G13 as a means of quantifying the nature of stellar orbits within Solar Neighbourhood--like volumes in the {\it Aquarius} simulations. However, the local stream density around each stellar particle in G13 was computed by following the neighbouring stellar particles that, at the corresponding stream's formation time, were located within spheres of radii larger than or equal to $4$ kpc radius, much larger than the kernel size we use in Figure~\ref{fig:1}. We now explore whether the above result holds even when larger local volumes are considered for the initial distribution of test particles. 

For this experiment we consider a guiding stellar particle that is moving on a regular orbit. In Figure~\ref{fig:nosabe} we show the time evolution of the OFLI for test particles initially distributed over configuration space kernels of different sizes. In all cases the particles were distributed over
the same Gaussian kernel in velocity space, with $\sigma_{\rm v} = 1$ km/s. The left panel shows the results obtained for $\sigma_{\rm x} = 0.005$ kpc. Within this relatively small sphere, the evolution of the OFLI for most test particles still reflects the behaviour of the guiding particle. 

In the middle panel of Figure~\ref{fig:nosabe} we use a configuration space kernel with $\sigma_{\rm x} = 0.1$ kpc. Although the majority of test particles still show a regular behaviour in this case, we start to find a significant fraction that show a chaotic behaviour in the same integration period. 
The right--hand panel shows results for $\sigma_{\rm x} = 4$ kpc. It is clear from this panel that such an extended initial distribution of test particles does not accurately reflect the dynamical behaviour of the guiding particle. The results presented in G13 may overestimate the fraction of chaotic orbits within each Solar Neighbourhood--like volume.

\subsubsection{Connecting the OFLI to a measure of the local stream density}

We have shown that \emph{initially nearby} particles in phase space show similar time evolution of the OFLI, which implies a common dynamical behaviour. We will now explore whether the behaviour of the OFLI accurately reflects the time evolution of local density along a stream. For this purpose, we follow the evolution of the local density around a sticky stellar particle and a chaotic stellar particle for periods of 10 Gyr. We refer to these stellar particles as guiding particles. Both guiding particles are located in a Solar Neighbourhood--like sphere at $z=0$. Estimates of the chaos onset times for these guiding particles are $\sim 80$ and 3 Gyr, respectively. Note that the following results do not depend strongly on our specific choice of stellar particles.

As before, we distribute ensembles of test particles around each guiding particle. The test particles are initially distributed as explained in Section~\ref{subsubsec:bridge-connect}, with $\sigma_{\rm x} = 0.2$ pc and $\sigma_{\rm v} = 1$ km/s. However, to accurately track the time evolution of the local density around both guiding particles for periods of 10 Gyr, a larger number of test particles, $10^4$, is considered for each ensemble. 
Two different time intervals, separated by several  Hubble times ($\sim 113$ Gyr), are considered. The integration timestep is $0.1$ Myr. As expected from their chaos onset times, during the first time interval the sticky orbit behaves like a regular orbit while the chaotic orbit shows its true nature. During the second time interval the sticky orbit behaves like a chaotic orbit. To estimate the local density at every timestep, we count the number of test particles within a radius of $0.1$ kpc around the guiding particle, and also discard from further consideration any particles beyond a radius of $2$ kpc. The number of test particles within the 0.1 kpc sphere is then normalised by the initial number of test particles. We call this quantity the normalised number of neighbouring particles: $\mathrm{N}_{i}$. Here $i = {\rm S}, \rm{C}$ refers to the normalised densities associated with the sticky and chaotic orbits, respectively.

\begin{figure}
\begin{center}
\begin{tabular}{c}
\hspace{-5mm}\includegraphics[width=1\linewidth]{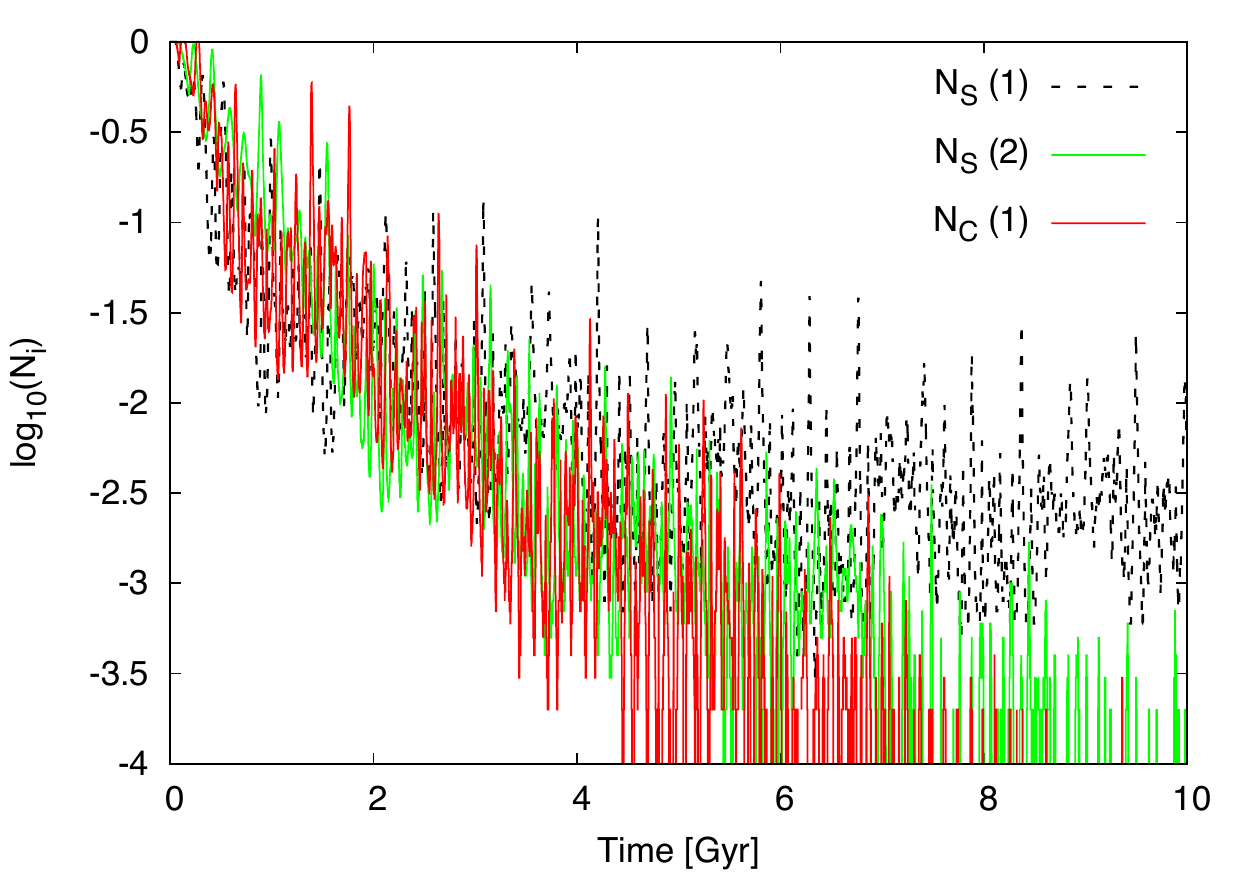}\\
\includegraphics[width=1\linewidth]{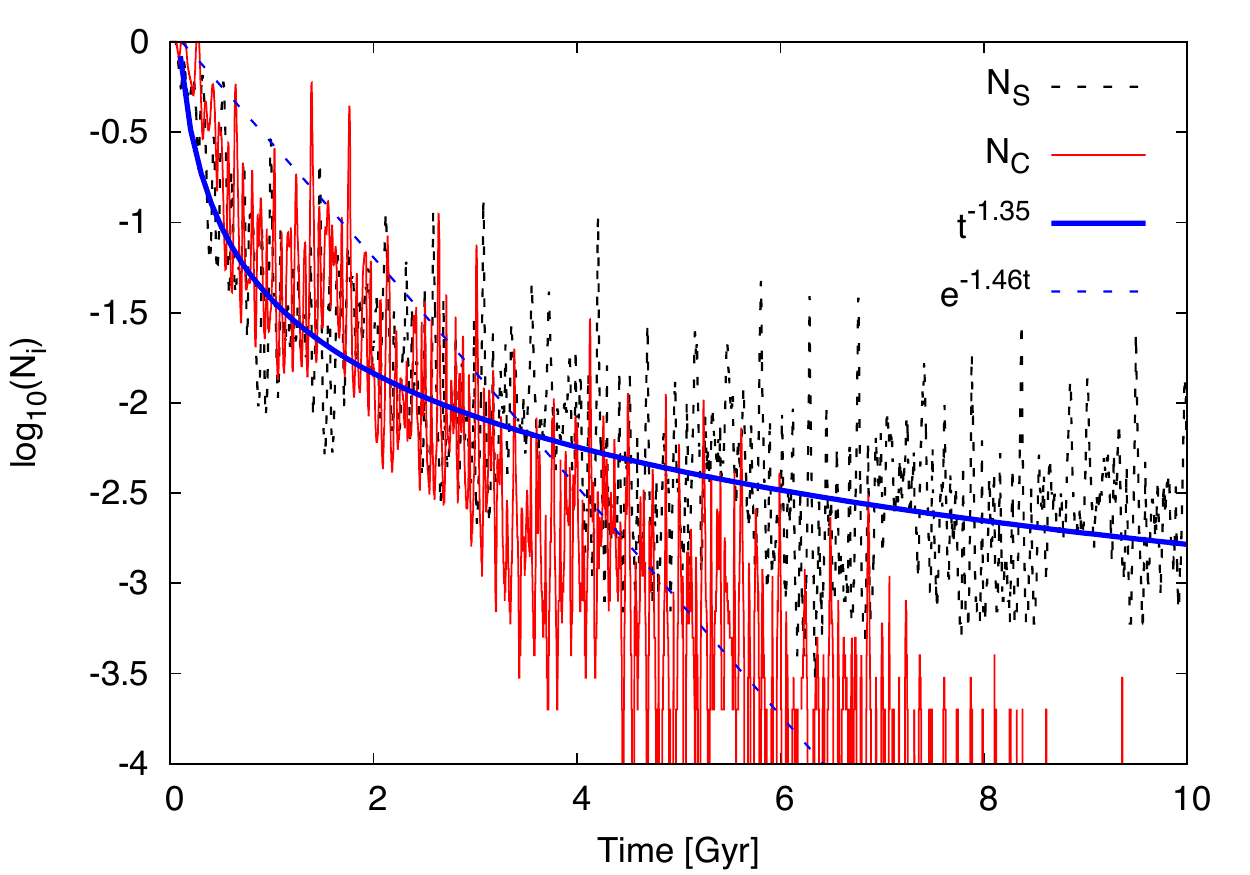}
\end{tabular}
\end{center}
\caption{Top panel: time evolution of the normalised number of neighbouring particles for the sticky orbit, $\mathrm{N}_{\mathrm{S}}$, over two non--consecutive 10 Gyr time intervals. The first interval (labelled 1) is taken when the guiding particle moves on a regular orbit and the second interval (labelled 2) is taken when the guiding particle moves on a chaotic orbit. We also show the time evolution of the normalised number of neighbouring particles for the chaotic orbit, $\mathrm{N}_{\mathrm{C}}$, over the first time interval. Bottom panel: a power law fit for $\mathrm{N}_{\mathrm{S}}$ and an exponential  fit for $\mathrm{N}_{\mathrm{C}}$ for the first interval only. Notice the logarithmic scale. The local density of stellar streams decreases with a power law function along a regular orbit and at an exponential rate along a chaotic orbit.}
\label{fig:2}
\end{figure}  

On the top panel of Fig.~\ref{fig:2} we present, with dark grey dashed and green solid lines, the time evolution of  $\mathrm{N}_{\mathrm{S}}$ for both time intervals (labelling the first 10 Gyr as interval 1 and the second 10 Gyr time interval as number 2). On the same panel we also present, with a red solid line, the time evolution of $\mathrm{N}_{\mathrm{C}}$ for the first 10 Gyr time interval. It is clear from this panel that the local density around the sticky orbit decreases in time significantly more slowly when the guiding particle's orbit is approximately regular. Notice that, in this regime, we are able to track the evolution of the local density for the full 10 Gyr period of integration. Conversely, in the chaotic regime, the number of neighbouring particles within 0.1 kpc, that have never been further than 2 kpc, becomes insufficient to track the local density after $\sim 6$ Gyr. Notice that the sticky particle and the chaotic particle show a similar evolution of their local densities during the chaotic regime. The more rapid decay of ${\rm {N_{\rm C}}}$ reflects the more chaotic nature of this orbit. On the bottom panel  of Fig.~\ref{fig:2} we show that a power law function describes well the time evolution of the local density around the sticky particle during the first 10 Gyr time interval; specifically ${\rm {N_{\rm S}}} \propto t^{-1.35}$. Conversely, an exponential relation is required to describe the time evolution of the local density around the chaotic particle $\mathrm{N}_{\mathrm{C}} \propto e^{-1.46 t}$.

This analysis demonstrates a very strong connection between the evolution of local (stream) density around a given guiding particle and the characterisation of its orbit provided by the OFLI. If the OFLI shows a linear growth, indicating regular behaviour, then the local density is expected to decrease as a power law with index less than or equal to $3$. Likewise, an exponential growth of the OFLI reflects an exponential decay of the corresponding local density. 

\subsubsection{The distribution of chaos onset times within Solar Neighbourhood--like volumes}
\label{subsubsec:relevance-time}

In the previous section we have shown that the OFLI is a powerful tool to characterise the time evolution of the local (stream) density around any stellar particle in our simulations. We will now use this indicator to quantify robustly the fraction of stellar particles within Solar Neighbourhood--like volumes that are moving on regular, sticky and chaotic orbits. If chaotic orbits are common in a phase space volume local to the Sun, then many phase space substructures (arising for example from past accretion events) may have been erased, due to the much shorter mixing timescales associated with such orbits.

In this experiment we will consider five different Solar Neighbourhood--like volumes, each extracted from a different {\it Aquarius} stellar halo. In all cases the spheres are centred at 8 kpc from the galactic centre and have a radius of 2.5 kpc (see Section \ref{subsec:meth-init}). In order to compute the OFLI for each stellar particle within these spheres, we integrate the equations of motion (Eqs.~\eqref{eq:3}) together with the first variational equations (Eqs.~\eqref{eq:3.5}), assuming smooth triaxial NFW DM haloes (Eq.~\eqref{eq:nfw_tri}) with parameters as given in Table~\ref{table:aquarius}. The timestep of integration is $1$ Myr and the total integration time 1000 Gyr. We use such a long timespan (more than 100 times the likely age of the Milky Way) to identify very sticky orbits reliably. Finally, we compute the distribution of estimated chaos onset times to obtain the percentages of particles moving on sticky and chaotic orbits.

\begin{table*}
\centering
\begin{minipage}{138mm}
\centering
\caption{Dynamical distributions of particles at $z=0$ from the {\it Aquarius Project}. The first column labels the simulated DM haloes. From left to right, the columns give: the number of particles, $N^{\star}$; the number of particles which computation was stopped, $N^{\circ}$; the number of particles on regular orbits, $N^{\star}_{\rm R}$; the number of particles on sticky orbits, $N^{\star}_{\rm S}$; the number of particles on chaotic orbits, $N^{\star}_{\rm C}$; the number of particles with values of the Dy$_{ts}$ $\lesssim$ Dy$_{ts}^{\mathrm{max}}$, $N^{\star}_b$ and the medians of Dy$_{ts}$ for the sticky, $\mu^{\rm S}_{1/2}$, and the chaotic distributions, $\mu^{\rm C}_{1/2}$ in Gyr.}
\label{table:particles}
\begin{tabular}{@{}ccccccccc} \hline \hline Name & $N^{\star}$ & $N^{\circ}$ & $N^{\star}_{\rm R}$ & $N^{\star}_{\rm S}$ & $N^{\star}_{\rm C}$ & $N^{\star}_b$ & $\mu^{\rm S}_{1/2}$ & $\mu^{\rm C}_{1/2}$ \\
\hline Aq--A2 & 1388 & 12 & 437 (31.48\%) & 644 (46.4\%) & 307 (22.12\%) & 1310 (94.38\%)& 0.295 & 0.256 \\
Aq--B2 & 10385 & 355 & 3618 (34.84\%) & 5069 (48.81\%) & 1698 (16.35\%) & 10273 (98.92\%)& 0.417 & 0.396 \\
Aq--C2 & 2291 & 43 & 756 (31.62\%) & 1015 (46.63\%) & 520 (21.75\%) & 2275 (99.3\%)& 0.297 & 0.263 \\
Aq--D2 & 3745 & 108 & 1087 (29.03\%) & 2213 (59.09\%) & 445 (11.88\%) & 3679 (98.24\%)& 0.518 & 0.409 \\
Aq--E2 & 1877 & 80 & 589 (31.38\%) & 888 (47.31\%) & 400 (21.31\%) & 1853 (98.72\%)& 0.494 & 0.426 \\
\hline
\end{tabular}
\end{minipage}
\end{table*}

As previously mentioned in Section \ref{sec:meth-ind-bridge}, we consider as a sticky orbit any orbit that has a chaos onset time larger than 10 Gyr (a Hubble time, roughly speaking). This definition is arbitrary. However, it allow us to make a clear distinction between orbits that could show some degree of chaotic mixing within a physically meaningful timescale and those for which chaos is completely irrelevant. 

As described in Section~\ref{subsec:meth-init}, only accreted stellar particles are analysed in these experiments. Table~\ref{table:particles} lists the total number of stellar particles considered in each stellar halo, $N^{\star}$. This number is slightly different from the number of stellar particles quoted in G13 (see their Table 3). This is due to a small number of stellar particles, $N^{\circ}$, that needed some numerical readjustments during the integration -- the computation of these particles was stopped in order to keep the computing time of the integrator bounded. 

\begin{figure}
\begin{center}
\begin{tabular}{c}
\hspace{-5mm}\includegraphics[width=1\linewidth]{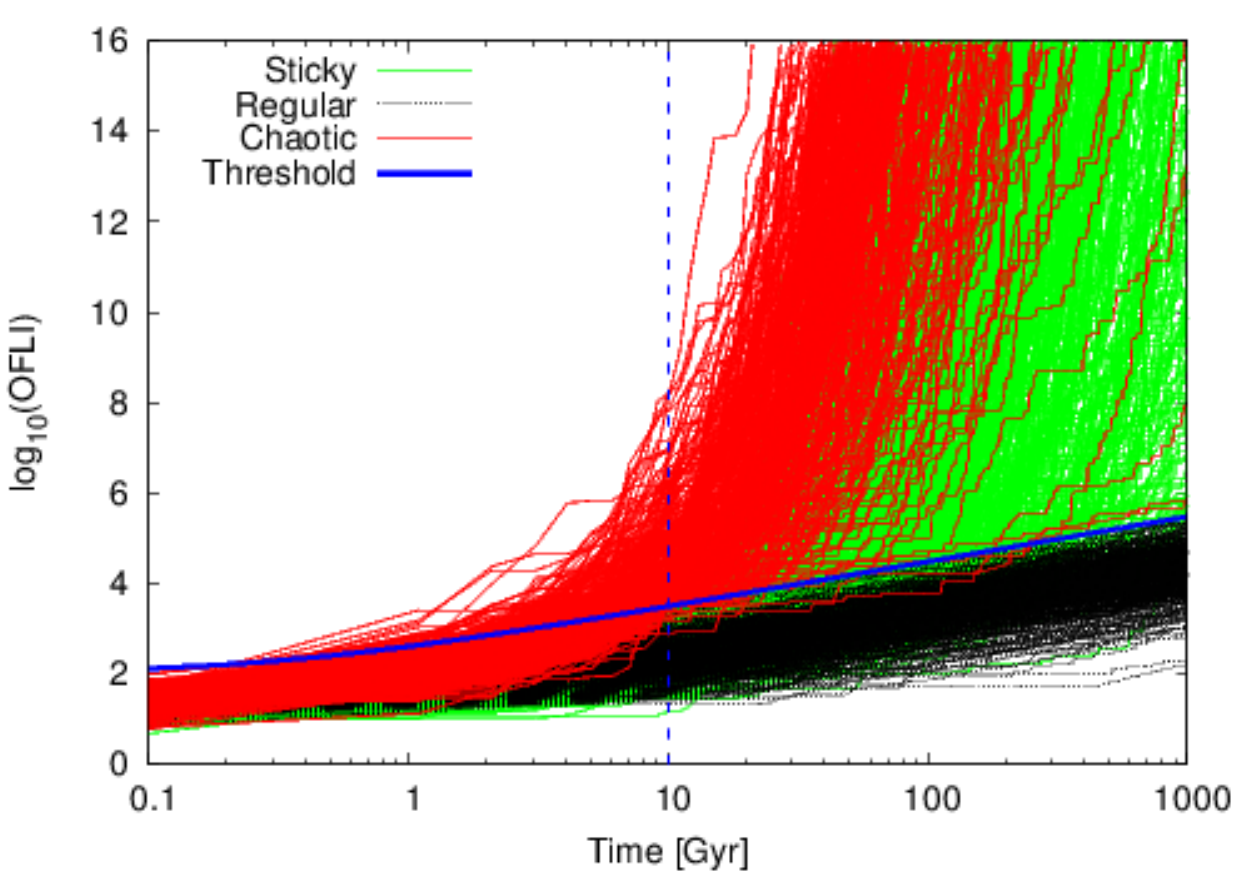}
\end{tabular}
\end{center}
\caption{Time evolution of the OFLI for the 1388 particles considered in the \emph{Aquarius Project} for the Aq--A2 DM halo and within an interval of time long enough to identify very sticky orbits (1000 Gyr). The upper limit used as a threshold for regular motion is depicted in solid blue. The 10 Gyr threshold is depicted with a vertical dashed--blue line. Notice the logarithmic scale. The three orbital components, i.e. the sticky, the regular and the chaotic components, are clearly distinguished by using the OFLI with both simple thresholds.}
\label{fig:3}
\end{figure}

The orbits of the particles selected for further study were then classified according to the shape of
of their OFLI time evolution curve. Individually inspecting each curve to estimate the chaos onset time is unfeasible for large samples of orbits. We therefore introduce a threshold OFLI value based on an upper limit for the typical linear behaviour seen for regular orbits. This threshold evolves linearly with time and envelopes all the curves that present a linear behaviour. On Figure~\ref{fig:3} we show an example of this procedure. The different lines show the time evolution of the OFLI for 1388 particles located within the Aq--A2 Solar Neighbourhood--like sphere. The threshold is indicated with a blue solid line. Every time the OFLI of any given stellar particle crosses this threshold, the corresponding particle is classified as either sticky  or chaotic. We reject all threshold crossings in the first Gyr of evolution, because this period can be considered as a transient stage of the indicator. Chaotic orbits are defined by threshold crossing within the first 10 Gyr of their evolution. This 10 Gyr \emph{barrier} is depicted in Fig.~\ref{fig:3} by a vertical dashed blue line. We denote the threshold crossing times of each stellar particle by T$^t_c$ (our estimation of the chaos onset time). The different colour--coded curves in Figure~\ref{fig:3} show the results of this classification applied to halo Aq--A2. Similar results are found for the other four haloes, which we do not show for the sake of brevity. Note that, in a very small number of cases, the OFLI of an orbit crosses the threshold early on, but later continues to evolve linearly with time. As a consequence, the number of chaotic orbits found within each volume may be slightly overestimated.

In Table~\ref{table:particles} we summarise the result of this experiment. It is interesting to see that, even in significantly triaxial and cuspy potentials, a significant fraction of orbits are regular even after 1000 Gyr of evolution. In all haloes, we find that $\approx 30\%$ of the stellar particles are moving on regular orbits. It also striking to find that, in all haloes, the fraction of stellar particles on sticky orbits is larger than $45\%$. As a consequence we find that only $\lesssim 20\%$ of orbits could be experiencing some degree of chaotic mixing, regardless of the scale and shape of the halo.

Stellar particles living in the innermost regions of each stellar halo are likely to be more bound and to have shorter dynamical timescales than those populating the outer galactic regions. It is interesting to study whether the distribution of orbital dynamical timescales (hereinafter Dy$_{ts}$) plays an important role in separating sticky from chaotic orbits. If chaotic orbits are preferentially found in the innermost galactic regions, then we may have been able to identify them simply because their corresponding Dy$_{ts}$ are small enough to reveal their nature in short integration times. On the other hand, for orbits probing the outer regions of the halo, with larger values of Dy$_{ts}$, very long integration times could be required for an accurate characterisation. It is thus important to understand whether or not our quantification of regular, sticky and chaotic orbits is biased by differences in the relative distributions of their orbital timescales.

\begin{figure}
\begin{center}
\begin{tabular}{c}
\hspace{-5mm}\includegraphics[width=1\linewidth]{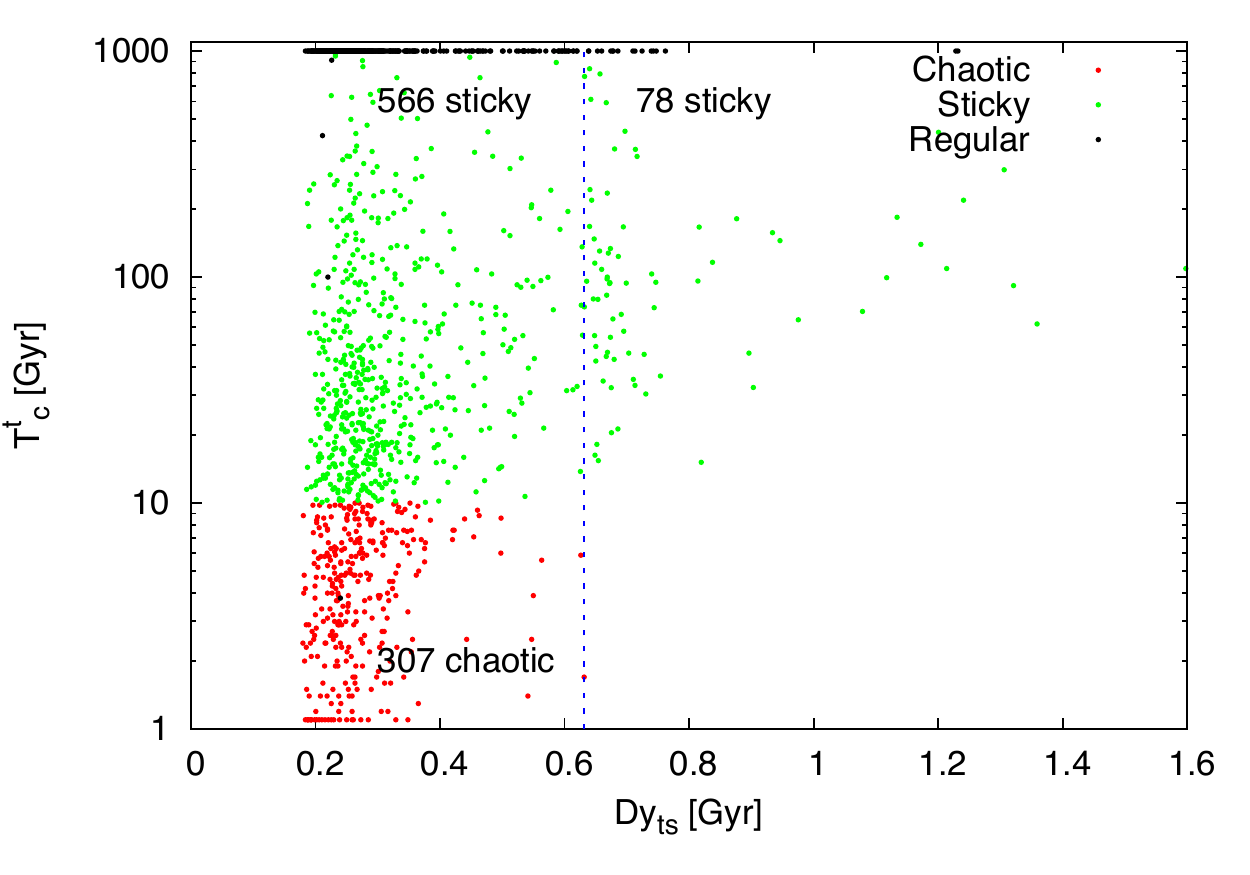}
\end{tabular}
\end{center}
\caption{The threshold crossing time, T$^t_c$, as a function of the dynamical timescale, Dy$_{ts}$. The chaotic orbits are depicted in red, the sticky orbits in green and the regular orbits in black as throughout the paper. The vertical dashed blue line is the threshold for Dy$_{ts}^{\mathrm{max}}$, i.e. $0.631$ Gyr for the halo Aq--A2. The distributions for both chaotic and sticky orbits are very similar.}
\label{fig:4}
\end{figure}  

To explore this we define a characteristic Dy$_{ts}$ as the time for a particle on a given orbit to undergo \textit{two} changes of the sign of its velocity component along the major axis of the stellar halo. In Figure~\ref{fig:4} we present, for halo Aq--A2, the distribution of Dy$_{ts}$ as a function of the estimated chaos onset time, T$^t_c$. Regular, sticky and chaotic orbits are depicted in black, green and red, respectively. As before, this classification is based on their values of T$^t_c$. It is very clear from this panel that chaotic and sticky orbits are not strongly segregated in Dy$_{ts}$. Note that many sticky orbits with T$^t_c$ $\geq 100$ Gyr can be found with values of Dy$_{ts}$ as small as $\sim 0.2$ Gyr. A similar result is observed for regular orbits.  Even though sticky orbits show a tail towards large values of Dy$_{ts}$, the populations of chaotic and sticky orbits seems to be characterised by similar distributions. This can be seen from the medians of the two distributions, given in Table~\ref{table:particles}. A vertical dashed--blue line indicates the location of the chaotic orbit with the highest value of Dy$_{ts}$: Dy$_{ts}^{\mathrm{max}}$, which is $\approx 0.631$ Gyr for halo Aq--A2. In Table~\ref{table:particles} we present the fraction, $N^{\star}_b$, of orbits (of all classes) with values of Dy$_{ts}$ less than the corresponding Dy$_{ts}^{\mathrm{max}}$. In all haloes,  $N^{\star}_b\gtrsim 95$\%. In addition, we find that only $\approx 12$\% of the sticky orbits are above this threshold in halo Aq--A2. This number is reduced to  $\lesssim 2$\% for the other haloes. This means that short orbital timescales are not the dominant factor that distinguishes between sticky and chaotic motion for stellar halo particles in our Solar Neighbourhood--like volumes. Instead, the results of this simple analysis demonstrate that the key distinction is the dynamical properties of the surrounding phase space volume.

So far, we have shown that a small but non--negligible fraction of orbits in Solar Neighbourhood--like volumes could indeed exhibit chaotic mixing. In what follows we will discuss the extent to which such mixing can erase nearby signatures of early stellar accretion onto the galaxy, within physically relevant periods of time.

\subsection{Global Dynamics and Diffusion}
\label{subsec:diffusion}
\subsubsection{Basic concepts}\label{subsubsec:form}

In this section we discuss a mechanism that could lead to global
chaotic mixing. In terms of the orbits of stars,  roughly 
speaking, chaotic mixing means that all trajectories starting in a small neighbourhood of a given point
in phase space (or on an energy surface) loose the memory of their initial conditions with time and
eventually appear uncorrelated. For $N$--dimensional systems ($N\ge 3$), since
KAM tori no longer divide the energy surface (see Appendix~\ref{sec:sticky} for a further description and references), it has been conjectured that any orbit lying in a thin chaotic layer around any resonance might visit the whole so--called Arnold web \citep[][]{A64,Ch79}.

In his note, Arnold proved the existence of motion along the stochastic layer of a given resonance in a rigorous way, for a rather simple near--integrable Hamiltonian.  He demonstrated that, for a very small perturbation it is possible to find a trajectory in the vicinity of the separatrix of a particular resonance that connects two points separated by an arbitrarily large distance, i.e. independent of the size of the perturbation, on a very long timescale.  Arnold's proof rests on the existence of a chain of hyperbolic tori along this resonance that may provide a path for the orbit -- if these tori are very close to each other, an orbit could transit over that chain.  Since every torus in the chain is labelled by an action value or unperturbed integral, a large but finite variation of this action could take place.  This mechanism, which permits motion along the resonance stochastic layer, is known (in the mathematical literature) as the {\it Arnold Mechanism}, while the term {\it Arnold diffusion} generally refers (in the physical literature) to a global phase space instability \citep[for details see][]{G90,L99,C02}, that is any (chaotic) orbit could visit the full Arnold web in a finite time.  The problem of how to extend the Arnold mechanism to a generic Hamiltonian remains unsolved. One of the main difficulties is related to the construction of such a chain of tori.

Regardless this severe limitation to understand Arnold diffusion as a global instability, it is assumed (in the physical literature) that Arnold diffusion does occur, and is responsible for chaotic mixing \citep[see for instance the discussion given in the last section of][] {C06}. For instance, assuming that in the phase space of steady state galaxies the chaotic component is a well--connected region (through some type of diffusion), \citet{M99} extended the classical Jeans theorem, which was formulated only for regular non-–resonant orbits \citep{BT87}, to take into account chaotic motion. However, as we will show in this section, chaotic diffusion does not play any significant role in connecting the whole chaotic component.  

In spite of the mathematical difficulties in dealing with this conjecture as a global property, a local formulation shows that chaos needs to be considered in the limit when $t\to\infty$ in order to observe any significant variation of the unperturbed integrals. This suggests that (strict) Arnold diffusion is irrelevant in the real world\footnote{Further discussion about this instability and the connection between the mathematical and physical approach can be found in \citet{GL13,CEGM14}.} \citep{ChV93,C02}.

In those real systems exhibiting a divided phase space, where the chaotic component is relevant (i.e. has a positive non--negligible measure) the timescale for any diffusion (not Arnold diffusion) would be much shorter but still very long \citep[see for instance][]{ChV97,GC04}. 

In the particular model considered here, we claim that 
chaotic mixing is almost irrelevant on cosmological timescales. Although this is the aim of a forthcoming paper, we would like to show,  
by simple arguments and computations,  that the so--called chaotic diffusion does not work for the model here consider.

\subsubsection{Analytic description}

We have already shown that $\Phi_{\rm TRI}$ can be approximated by
Eq.~\eqref{phi_sphe} (see Section \ref{subsec:meth-pot}), which for simplicity we recast as

\begin{multline}
\Phi_{\rm TRI}\approx\Phi_0(r)+\Phi_1(r)[\mu_1\cos2\varphi+\mu_2\cos2\vartheta+\\+\mu_3\cos2(\vartheta\pm\varphi)],
\nonumber
\end{multline}
where $\mu_s=\varepsilon_2-\varepsilon_1,\, -\varepsilon_1,\, \varepsilon_1/2$,
and since  $\varepsilon_1$ and $\varepsilon_2$ are assumed to be small parameters, $\mu_s\ll 1$. 

Therefore, the Hamiltonian takes the form:
\begin{equation*}
\mathcal{H}(\mathbf{p},\mathbf{r})=\mathcal{H}_0(\mathbf{p},r,\vartheta)+\hat{\Phi}_1(\mathbf{r}),
\end{equation*}
with
\begin{equation*}
\mathcal{H}_0(\mathbf{p},r,\theta)=\frac{p_r^2}{2}+\frac{p_{\vartheta}^2}{2r^2}+\frac{p_{\varphi}^2}{2r^2\sin^2\vartheta}+\Phi_0(r),
\end{equation*}
and
\begin{equation*}
\hat{\Phi}_1(\mathbf{r})= \Phi_1(r)\left[\mu_1\cos2\varphi+\mu_2\cos2\vartheta+\mu_3\cos2(\vartheta\pm\varphi)\right],
\end{equation*}
where
\begin{equation*}
p_r=\dot{r},\qquad p_{\vartheta}=r^2\dot{\vartheta},\qquad
p_{\varphi}=r^2\dot{\varphi}\sin^2\vartheta.
\end{equation*}
In fact, $\mathcal{H}_0$ is an integrable Hamiltonian being
\begin{equation*}
\mathcal{H}_0=\mathrm{E}_0,\qquad
\mathrm{L}_z= p_{\varphi},\qquad
\mathrm{L}^2=p_{\vartheta}^2+p_{\varphi}^2\csc^2\vartheta,
\end{equation*}
the three global unperturbed integrals, while
$\hat{\Phi}_1$ can be considered as a \emph{small} perturbation.
The dependence of $\hat{\Phi}_1$ on $(\vartheta, \varphi)$ leads to 
variations of the angular momentum and its components. Indeed,
\begin{eqnarray*}
\frac{dL_{z}}{dt}&=&[\mathrm{L}_z,\mathcal{H}]=
-\frac{\partial\hat{\Phi}_1}{\partial\varphi},\\
\frac{d\mathrm{L}^2}{dt}&=&[\mathrm{L}^2,\mathcal{H}]=
-2p_{\vartheta}\frac{\partial     \hat{\Phi}_1}{\partial\vartheta}-\frac{2p_{\varphi}}{\sin^2\vartheta}\frac{\partial \hat{\Phi}_1}{\partial\varphi},
\end{eqnarray*}
which are of order $\mu_s$ and therefore assumed to be small.

We are aware that the assumption $\mu_s\sim \varepsilon_1, \varepsilon_2\ll 1$ 
holds only marginally for any of the DM halo models considered here. 
However, the approach described above provides an appropriate physical insight into the problem.

\subsubsection{Numerical experiments}
\label{subsub:diffusion-numerical}

In order to perform numerical experiments on a given constant energy
surface on the $(\mathrm{L}^2,\mathrm{L}_z)$ plane that mimics our Solar Neighbourhood,
we consider the triaxial extension of the NFW model (Eq.~\eqref{eq:nfw_tri}) with parameters corresponding to the Aq--A2 DM halo. Initial conditions for an ensemble of test particles are obtained as follows. 

\begin{figure}
\begin{center}
\hspace{-5mm}\includegraphics[width=1\linewidth]{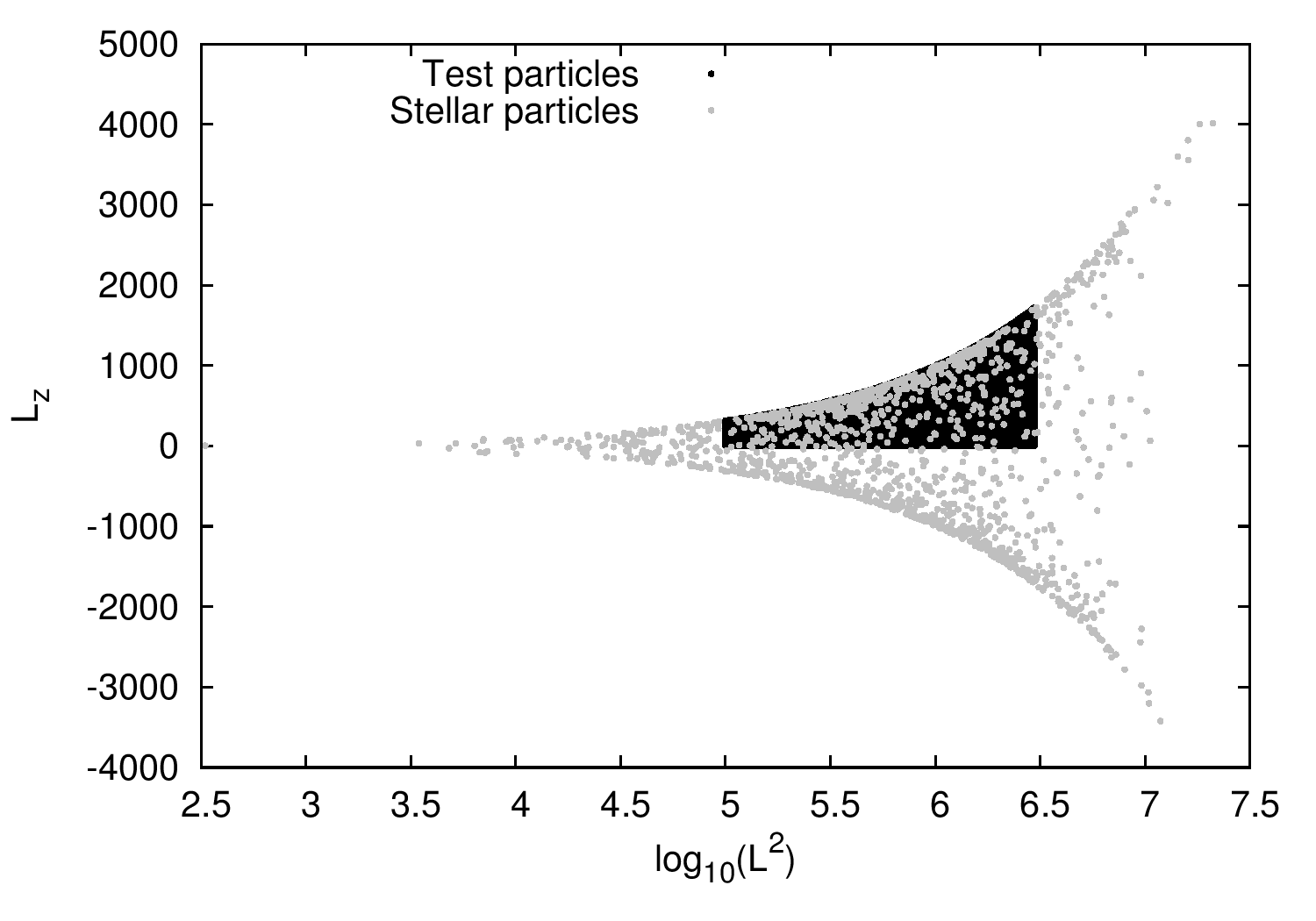}
\end{center}
\caption{Ranges in $\mathrm{L}^2$ (logarithmic scale) and $\mathrm{L}_z$
for the whole set of 1400 particles (grey) of the
Aq--A2 DM halo. In black, 
the region of the plane to be considered in the experiments.}
\label{fig:angular-momentum}
\end{figure}

First, we fix  $(x_0,y_0,z_0)=(8,0,0)$ kpc (i.e. the position of the {\it Sun}), and adopt the mean value of the energy distribution of the stellar particles located within a 2.5 kpc sphere,  $\langle \mathrm{E}\rangle = \mathrm{E}_0\simeq -204449$ km$^2$ s$^{-2}$. The remaining two phase space coordinates are obtained by sampling a region of the $(\mathrm{L}^2,\mathrm{L}_z)$ plane with a regular grid. The region of the $(\mathrm{L}^2,\mathrm{L}_z)$ plane  explored is determined such that, in both dimensions, $\approx 80\%$ of the corresponding stellar particles are enclosed. In Fig.~\ref{fig:angular-momentum} we show the selected region of the $(\mathrm{L}^2,\mathrm{L}_z)$ plane in which global dynamics will be displayed. We also include  the corresponding stellar particles with their actual values of $(\mathrm{L}^2,\mathrm{L}_z)$. \emph{We stress that, in the following experiments, the orbits of the test particles are calculated using the equations of motion and the first variational equations for the full triaxial NFW model}.

\begin{figure*}
\begin{center}
\begin{tabular}{cc}
\hspace{-5mm}\includegraphics[width=0.5\linewidth]{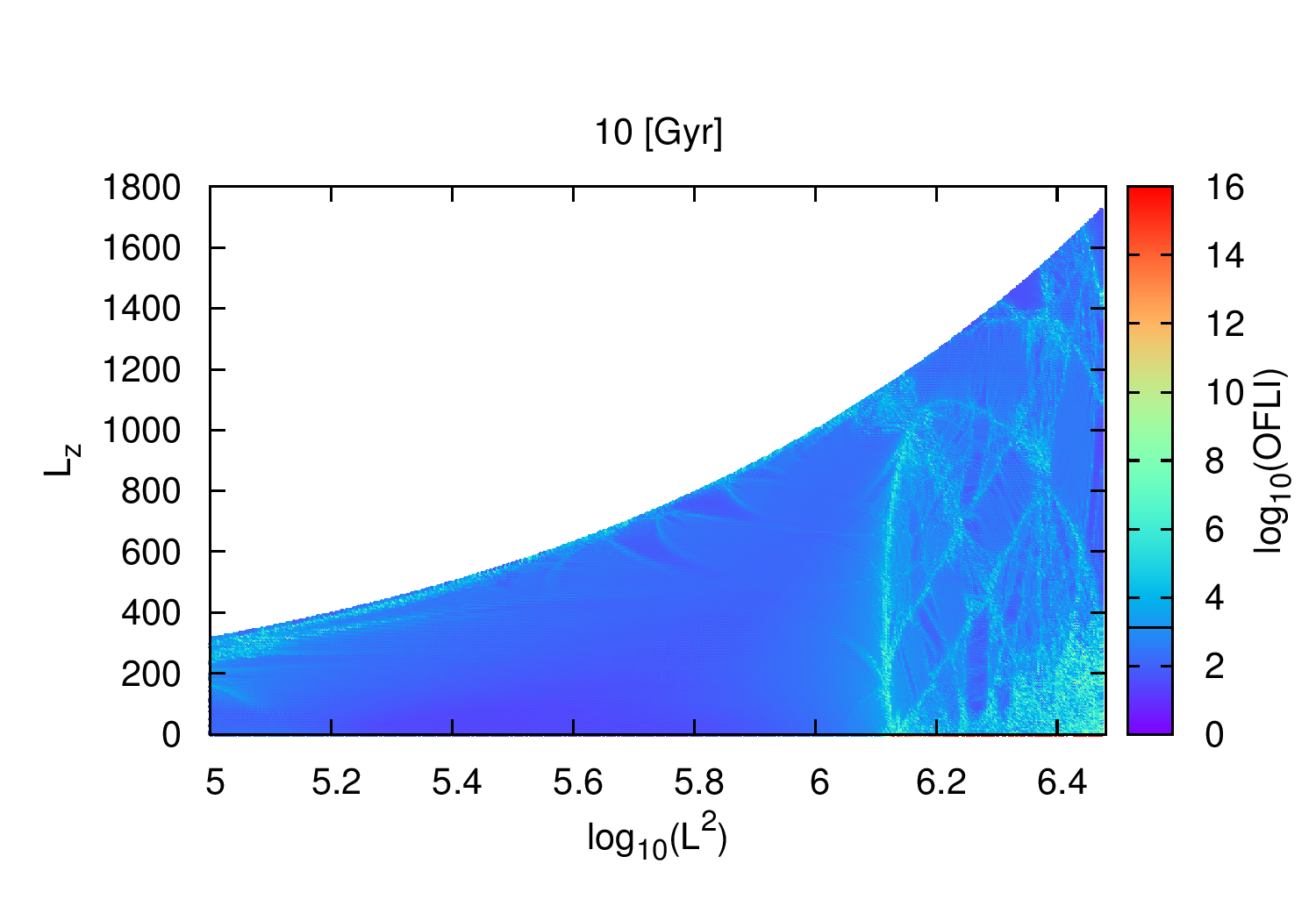}&
\includegraphics[width=0.5\linewidth]{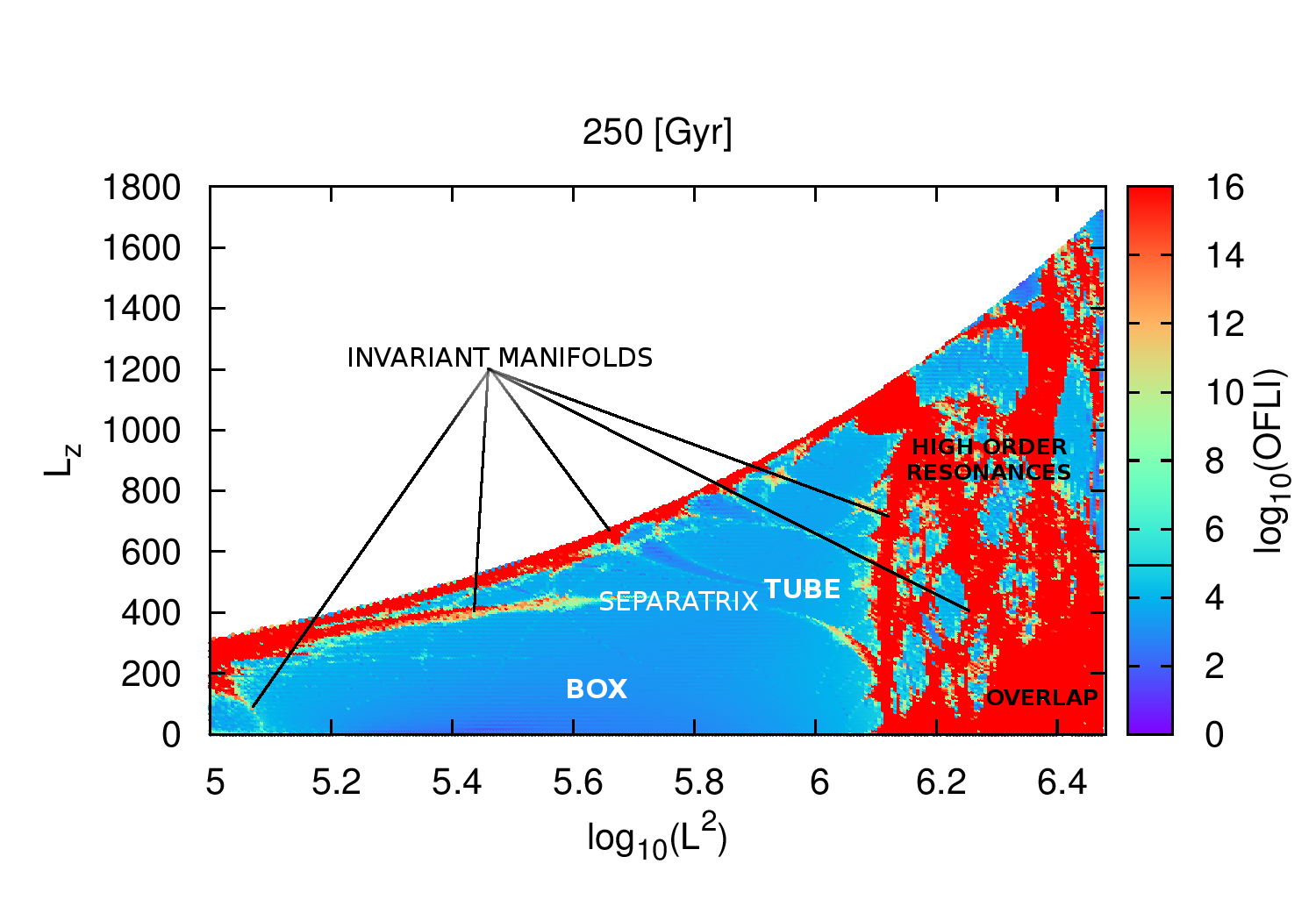}
\end{tabular}
\end{center}
\caption{OFLI contour plots for $10$ (left panel) and $250$ (right panel) Gyr for the Aq--A2 halo model for $(x_0, y_0,z_0)=(8,0,0),\;\mathrm{E}_0\simeq -204449$ km$^2$ s$^{-2}$. The solid black lines in the colour bars indicate the values of the threshold taken to distinguish regular from chaotic motion. Then, warm colours indicate chaotic motion while cool colours represent regular motion. The Arnold web is mostly unveiled and, as we can see from the warm colours of the right panel, it covers a considerable domain in phase space.}
\label{fig:404}
\end{figure*}

A global dynamical portrait of the system after an integration time of $10$ Gyr is shown in the left panel of Fig.~\ref{fig:404}, as a contour plot of the OFLI for a grid of $450\times450$ (yielding a total of $77967$ initial conditions). For this timescale,
almost the whole region of angular momentum space appears regular. Only a small number of invariant manifolds
and narrow resonances are observed. The invariant manifolds (separatrices) show up as arcs that separate different orbital families (the large resonance domains) while the small (or high--order) resonances arise as \emph{channels}, the centres of which correspond to a chain of stable 2D resonant elliptic tori and the \emph{margins} of which are formed by a chain of 2D hyperbolic (or unstable) tori. In any case, this figure shows that most of the phase space seems to be populated by regular orbits.

Although the perturbation is not actually small enough for the Aq--A2 halo model, namely, $\varepsilon_1\approx 0.17,\,\varepsilon_2\approx 0.63$, chaos seems
to be almost irrelevant after 10 Gyr of evolution. The amount of chaos
observed in this experiment, as measured in Section~\ref{subsubsec:relevance-time}, is $\simeq 12.44\%$. Consequently, no secular variation of the unperturbed integrals
$(\mathrm{L}^2,\mathrm{L}_z)$ is expected, since diffusion can only occur within
a connected chaotic region of finite size. The main effect of the perturbation
is to generate new families of orbits that, on timescale, are
regular box and tube orbits and subfamilies of the latter.

To look for diffusive phenomena or chaotic mixing in this model, we 
perform a second computation of the OFLI over a larger timescale of 250 Gyr. Although meaningless from a 
physical point of view, this experiment serves to unveil chaotic motion that still appears regular at 10 Gyr,
mainly as the result of sticky orbits. We are also interested in the timescale on which such diffusion takes 
place.

In the right panel of Fig.~\ref{fig:404}  we show  an OFLI contour plot for a grid of $300\times300$ ($34670$ initial conditions). This reveals some hyperbolic structures, shown in red. The true Arnold web is 
mostly unveiled -- it covers a considerable domain in phase space. Indeed, the fraction of chaos 
observed in this experiment amounts to the $\simeq 43.37\%$ of the orbits considered. Since a connected chaotic region of noticeable size exists, some secular variation of the unperturbed integrals $(\mathrm{L}^2,\mathrm{L}_z)$ would be expected, 
giving rise to \emph{fast} diffusion (as we discuss below). For this longer timescale,
the invariant manifold separating box from tube orbits in the more regular part of phase space (labelled as separatrix) is more clearly outlined. 
In addition, the high order resonant structure on the right, which appeared as channels crossed by several 
narrower resonances on the plot for $10$ Gyr, now shows up as an entangled assemblage of unstable manifolds, 
leading to strongly unstable or chaotic dynamics (red component in the right panel of Fig.~\ref{fig:404}).

\begin{table}
\centering
\caption{Ensembles of $90000$ initial conditions sampled uniformly in a box of size $\sim10^{-6}$, the  centre of which (given in the table) has been identified as corresponding to a chaotic orbit (recall that in the figures displaying the evolution of the unperturbed integrals, $\mathrm{L}^2$ is given on a logarithmic scale).}
\label{table:ens}
\begin{tabular}{@{}cll} \hline \hline Ensemble & $\log_{10}(\mathrm{L}^2)$ & $\mathrm{L}_z$\\\hline 
(i) &  5.045  & 275  \\
(ii) & 6.165 & 1000 \\
(iii) &  6.145 & 25 \\
(iv) &  6.405 & 25 \\
\hline
\end{tabular}
\end{table}

The long--term diffusion of the unperturbed integrals of this system are determined by the topology of all its resonances (the right panel of Fig.~\ref{fig:404}) which have been detected efficiently by our application of the OFLI. 
To gain insight into how this
diffusion operates, and to illustrate the roam of the unperturbed integrals, we have traced several orbits
with initial conditions embedded in different stochastic
domains in the $(\mathrm{L}^2,\mathrm{L}_z)$ plane. 
The wandering of the unperturbed integrals has been followed over $250$ Gyr for ensembles of $90000$ initial conditions sampled uniformly in boxes of size $\sim10^{-6}$. The centres of these boxes (listed in Table~\ref{table:ens}) are identified as the most chaotic regions by the OFLI. These ensembles are indicated by small green rectangles in the following figures.  
The equations of motion were integrated with a time--step of 1 Myr. 
Each crossing of an orbit through a spherical shell of radius $0.1$ kpc around the {\it Sun} is depicted as a black dot in the OFLI contour plot for the $250$ Gyr interval.

\begin{figure}
\begin{center}
\hspace{-5mm}\includegraphics[width=1\linewidth]{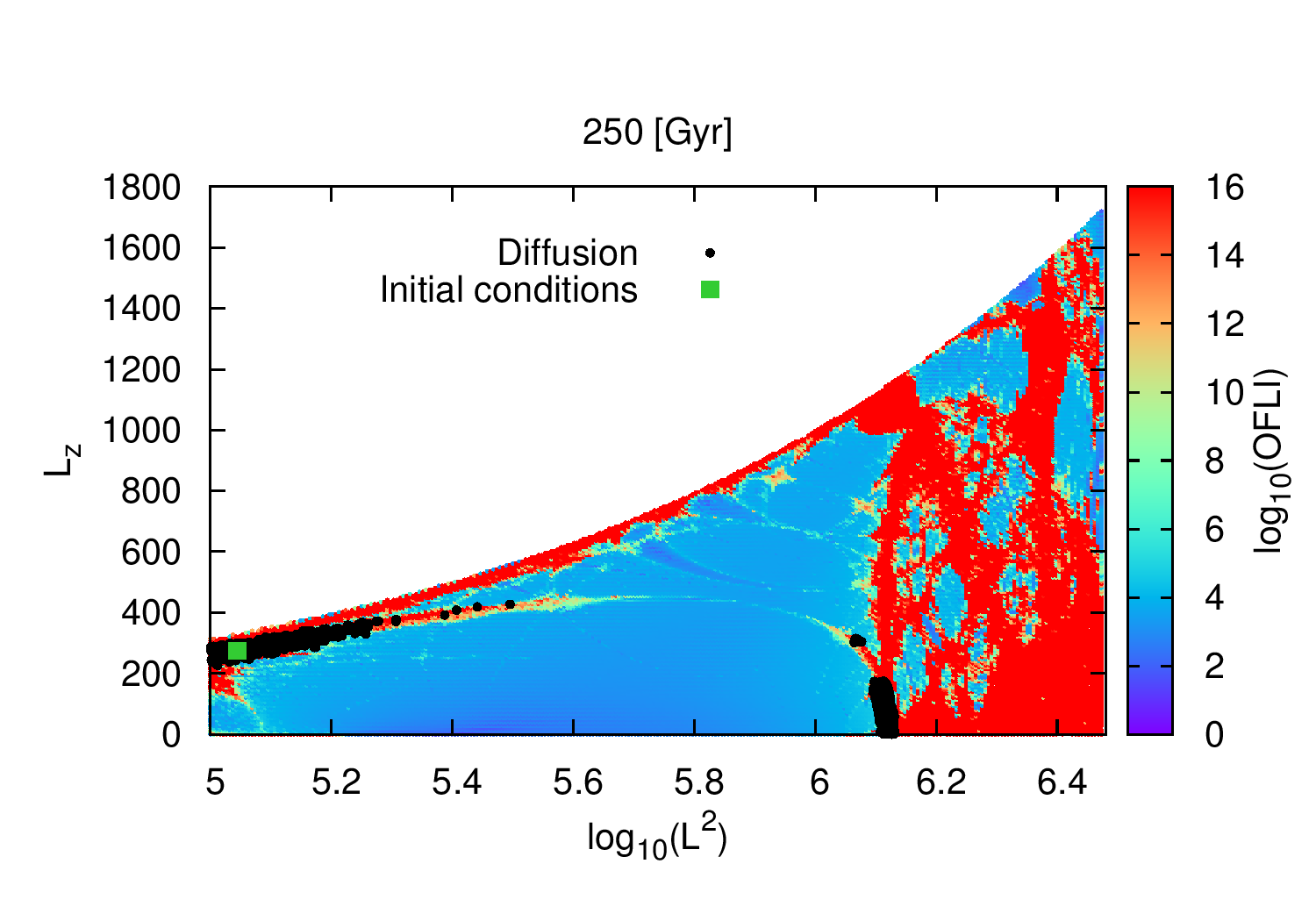}
\end{center}
\caption{Long--term diffusion over $250$ Gyr  
for ensemble (i) of initial conditions (depicted in green) overplotted on the Arnold web. Diffusion proceeds along the stochastic layer separating box from tube orbits.}
\label{fig:406}
\end{figure}

Fig.~\ref{fig:406} shows the time evolution of ensemble (i). Diffusion proceeds along the stochastic layer separating box from tube orbits. Recall that, even in the nearly completely regular scenario (for an almost
vanishing perturbation, i.e.  $\mu_s\to 0$),
the orbits will remain in an exponentially narrow
chaotic domain around the separatrix, at least for very large times.

\begin{figure}
\begin{center}
\hspace{-5mm}\includegraphics[width=1\linewidth]{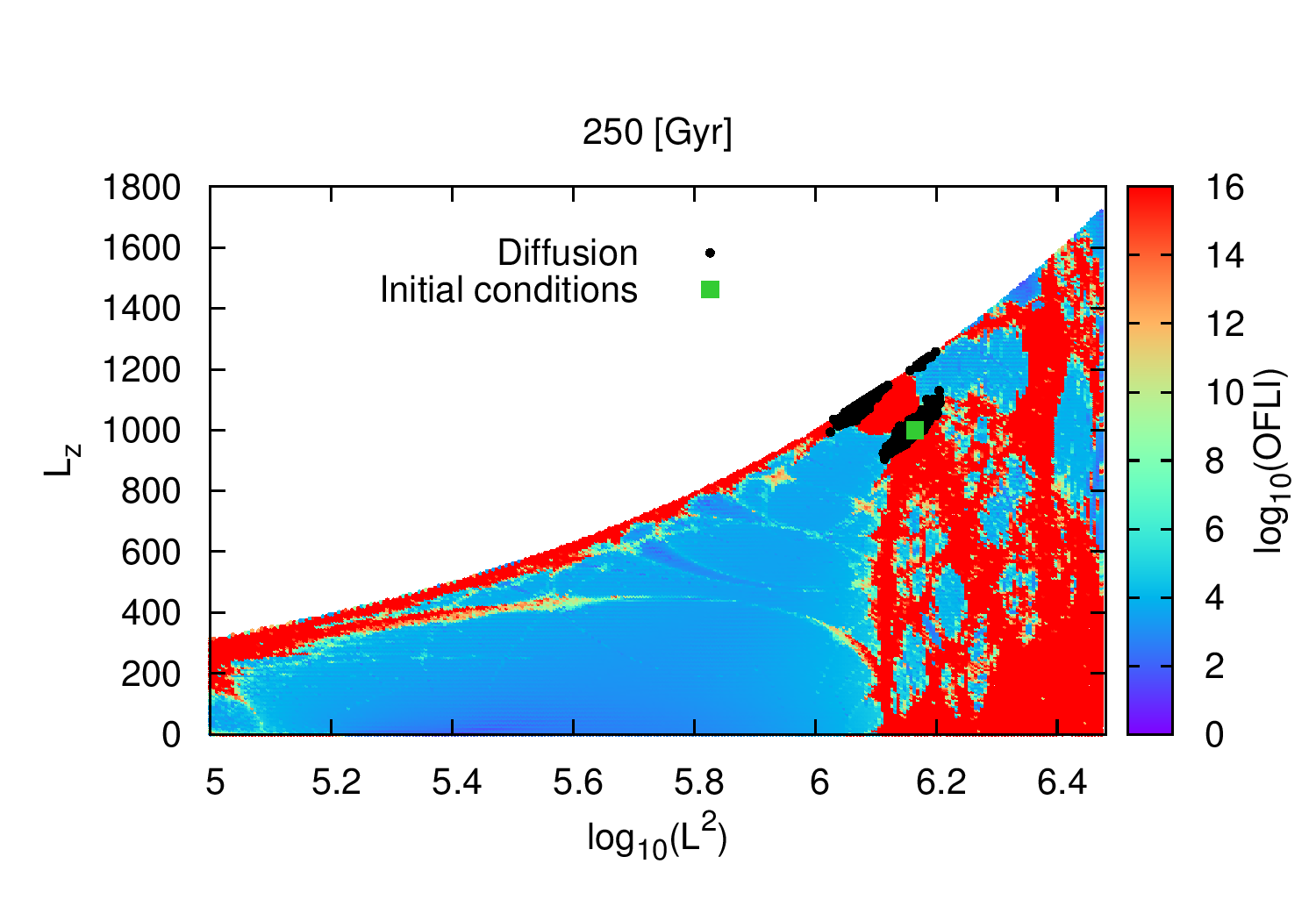}
\end{center}
\caption{Long--term diffusion over $250$ Gyr  
for ensemble (ii) of initial conditions (depicted in green) overplotted on the Arnold web. The unperturbed integrals remain confined to a rather small domain, hence diffusion turns out to be mostly inefficient.}
\label{fig:407}
\end{figure}

Meanwhile, when the evolution of ensemble (ii) is considered (Fig.~\ref{fig:407}), we recognise that the jump in pseudo--integrals between the two thin parallel curves corresponds to the borders of a resonance. In this case, the unperturbed integrals remain confined to a rather small domain, so that diffusion turns out to be inefficient for a rather large interval.

We stress that the diffusion we observe has some geometrical resemblance to the
theoretical conjecture of Arnold, according to which diffusion proceeds through phase space along the
chaotic layers of the full resonance web. However, it is clear that Arnold's
mechanism is not the way to understand this diffusion, since not only it is 
impossible to find any path for the orbits, but also the perturbation is not small enough ($\mu_s\lesssim 1$).
Thus the confined variation of $\mathrm{L}^2$ and $\mathrm{L}_z$ we find should be interpreted 
in terms of another regime, the well known \emph{overlap of resonances}. Even though
fast diffusion can occur in this scenario, clearly this is not the case here.

\begin{figure*}
\begin{center}
\begin{tabular}{cc}
\hspace{-5mm}\includegraphics[width=0.5\linewidth]{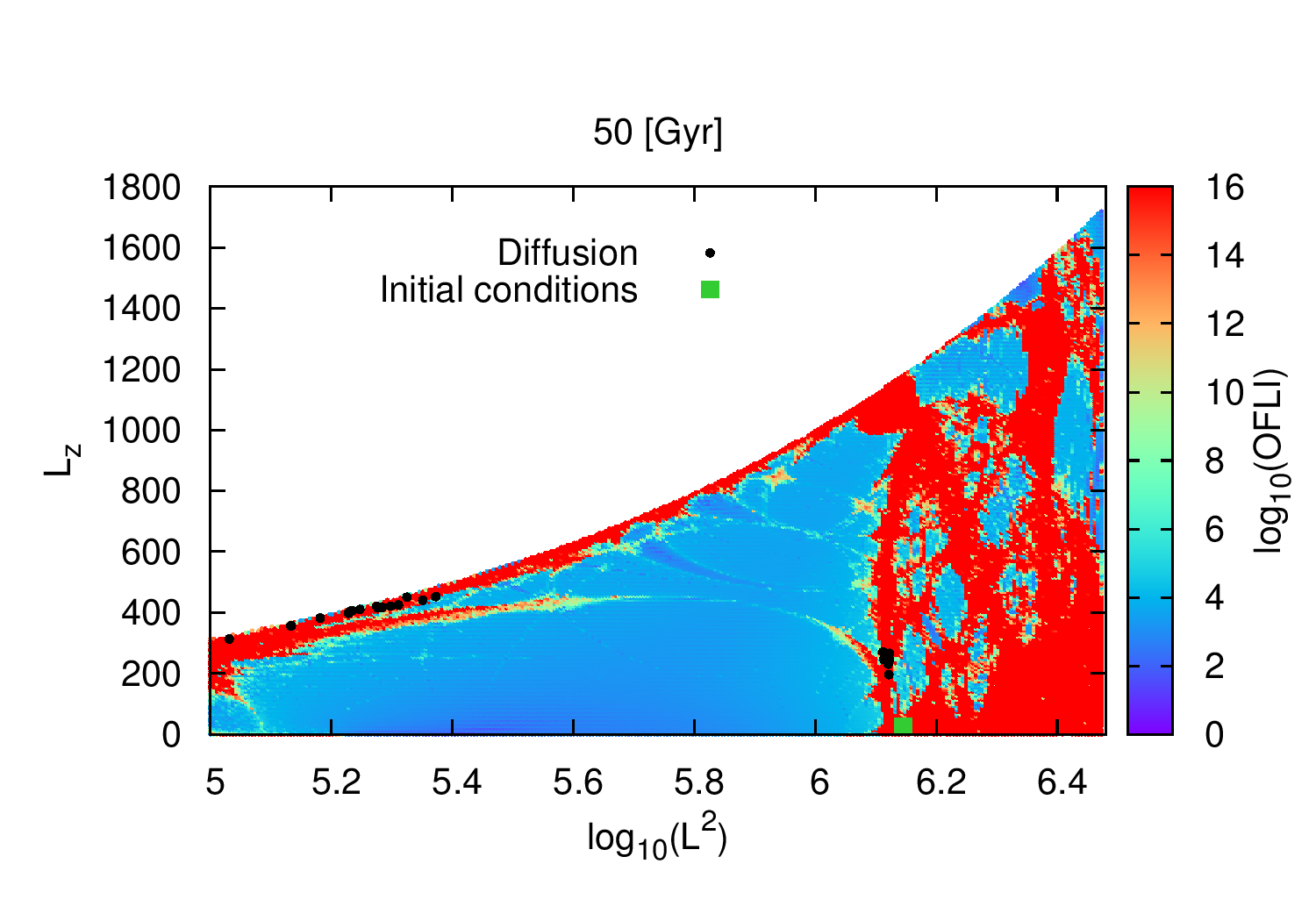}&
\includegraphics[width=0.5\linewidth]{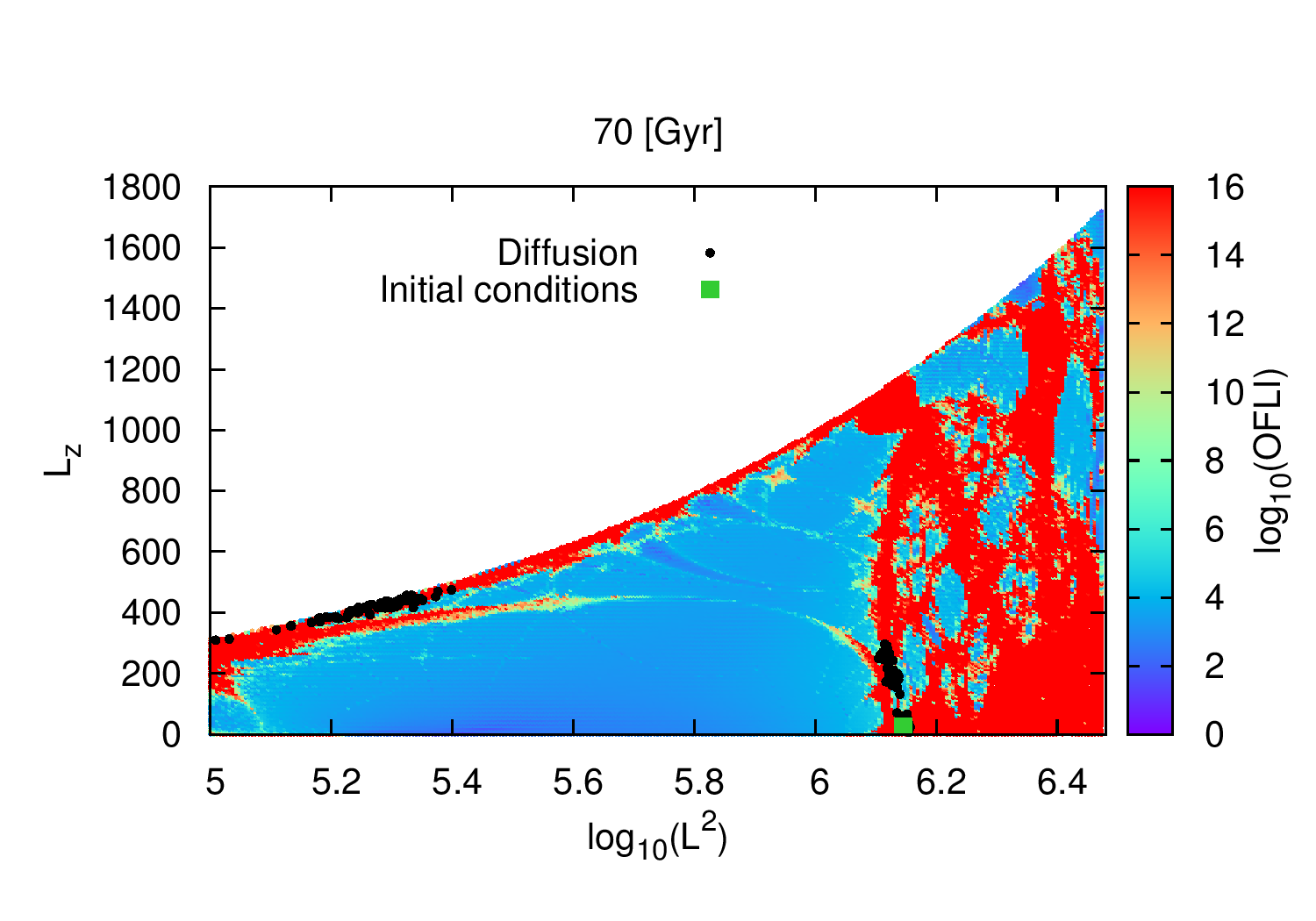}
\end{tabular}
\end{center}
\caption{Diffusion over $50$ Gyr (left panel) and $70$ Gyr (right panel) for ensemble (iii) of initial conditions (depicted in green) overplotted on the Arnold web as in previous figures. Diffusion advances along the outermost edge of the separatrix discriminating box from tube orbits.}
\label{fig:410}
\end{figure*}  

\begin{figure*}
\begin{center}
\begin{tabular}{cc}
\hspace{-5mm}\includegraphics[width=0.5\linewidth]{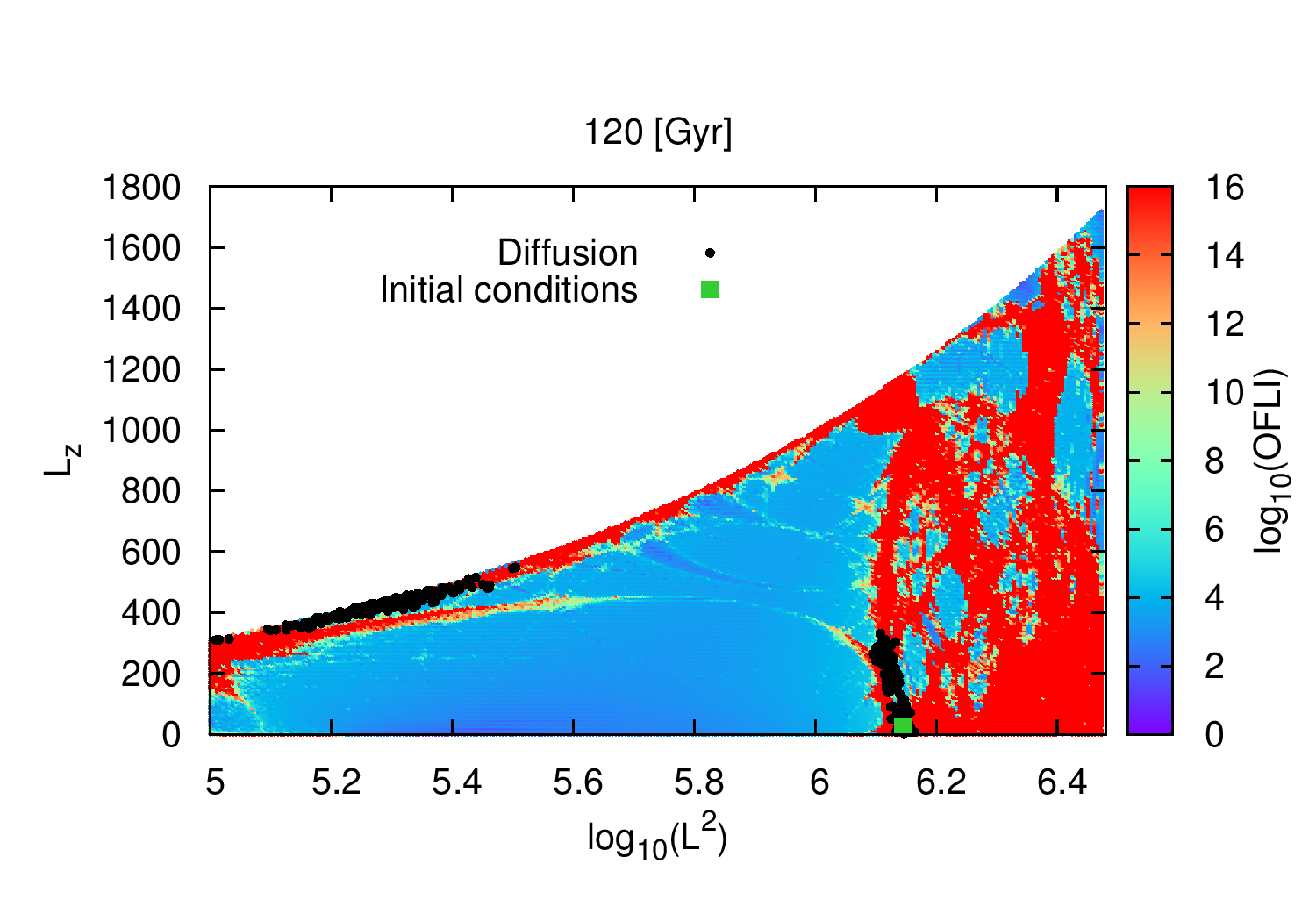}&
\includegraphics[width=0.5\linewidth]{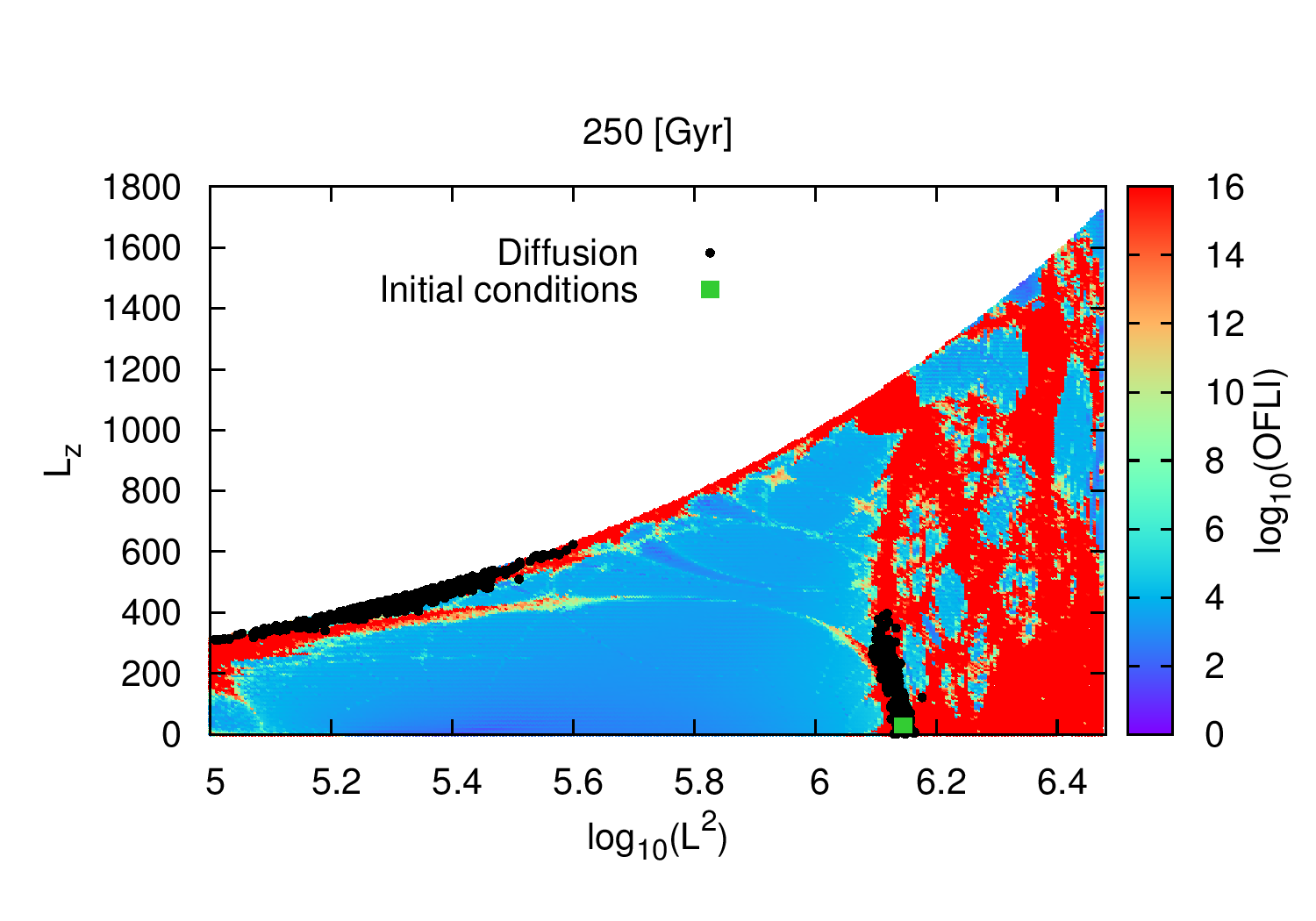}
\end{tabular}
\end{center}
\caption{Long--term diffusion over $120$ Gyr (left panel) and $250$ Gyr (right panel) for ensemble (iii) of initial conditions (depicted in green) overplotted on the Arnold web as in previous figures. Diffusion spreads out to cover the full width of the resonances.}
\label{fig:408}
\end{figure*}

To illustrate how diffusion progresses, Fig.~\ref{fig:410} shows snapshots corresponding to 50 (left panel) and 70 Gyr (right panel) for a third ensemble. We notice that diffusion advances along the outermost edge of the separatrix discriminating box from tube orbits, near the bottom of the figure, and climbs to slip over the 
left part of the web's upper border. When a larger time interval is considered, the points are seen to spread out to cover the full width of both resonances as shown in Fig.~\ref{fig:408} for 120 (left panel) and 250 Gyr (right panel). 

\begin{figure*}
\begin{center}
\begin{tabular}{cc}
\hspace{-5mm}\includegraphics[width=0.5\linewidth]{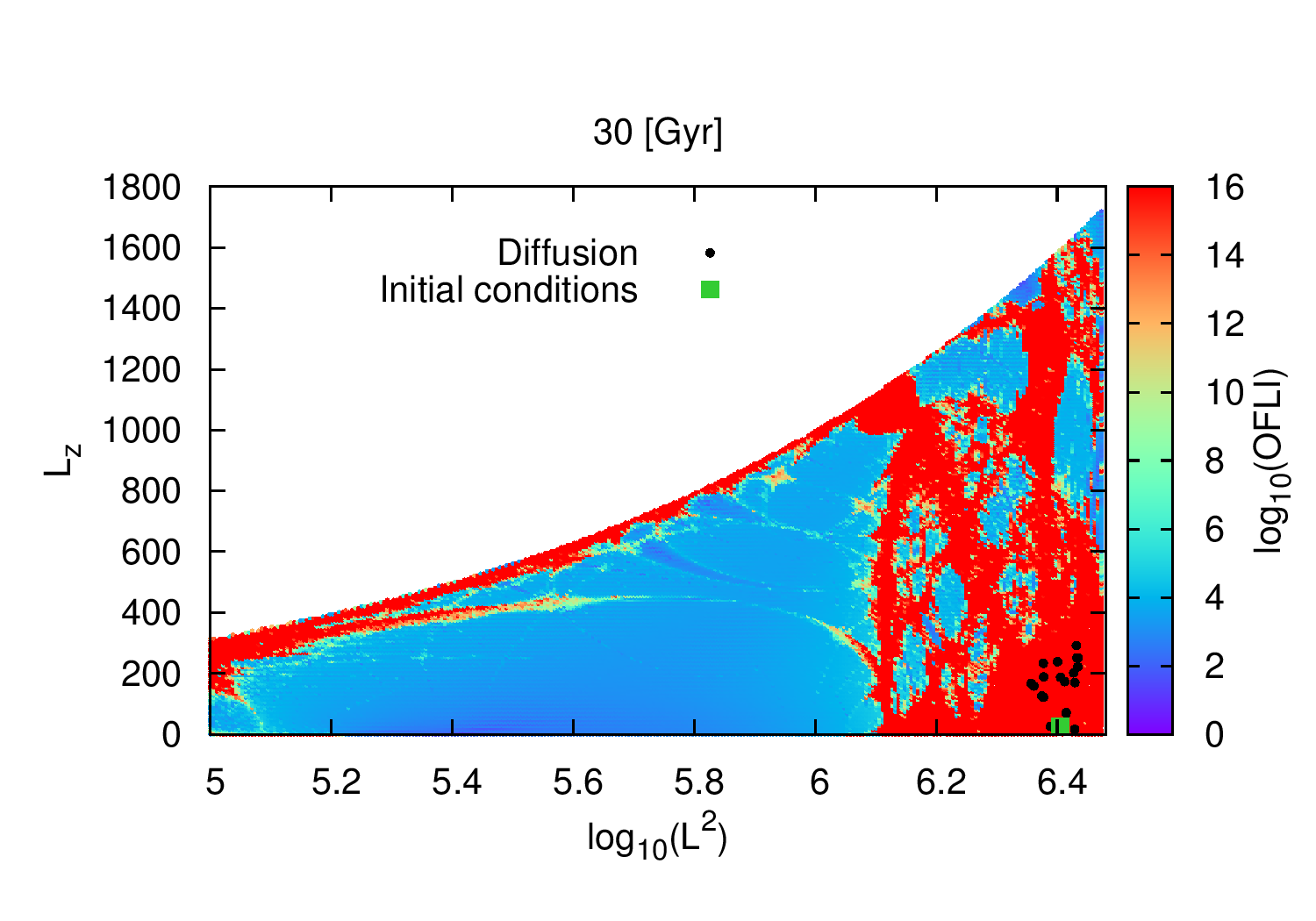}&
\includegraphics[width=0.5\linewidth]{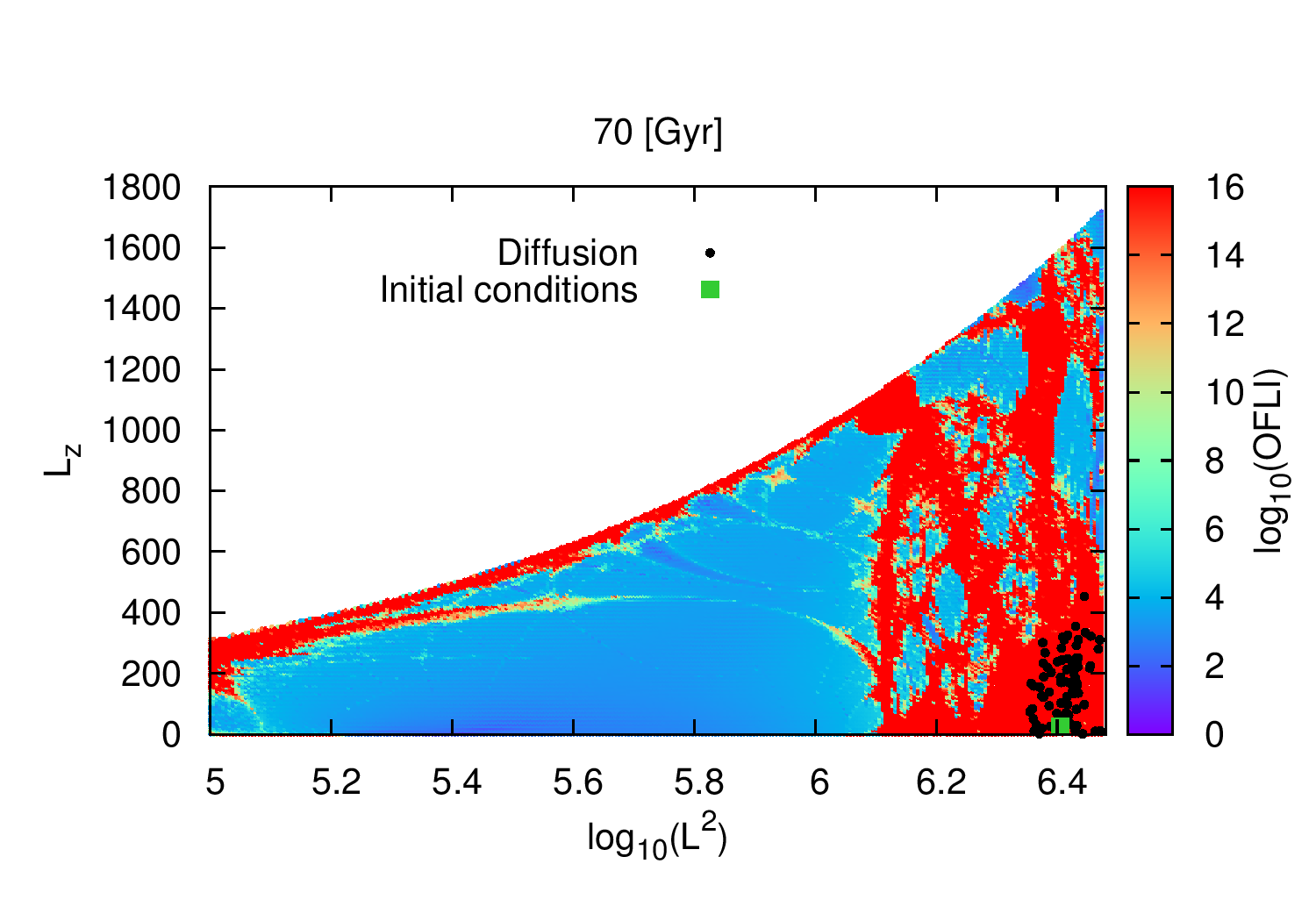}
\end{tabular}
\end{center}
\caption{Diffusion over $30$ Gyr (left panel) and $70$ Gyr (right panel) for ensemble (iv) of initial conditions (depicted in green) overplotted on the Arnold web as in previous figures. The unperturbed integrals wander over the small chaotic sea.}
\label{fig:411}
\end{figure*}

Meanwhile, for ensemble (iv) after 30 Gyr (Fig.~\ref{fig:411}, left panel) the unperturbed integrals wander over the small chaotic sea at the bottom right corner of the web, where an intricate overlap of resonances is observed.
For a larger timescale (70 Gyr, right panel) the roaming of ($\mathrm{L}^2$, $\mathrm{L}_z$) is confined to a domain of nearly the same extent (naturally more populated).

\begin{figure*}
\begin{center}
\begin{tabular}{cc}
\hspace{-5mm}\includegraphics[width=0.5\linewidth]{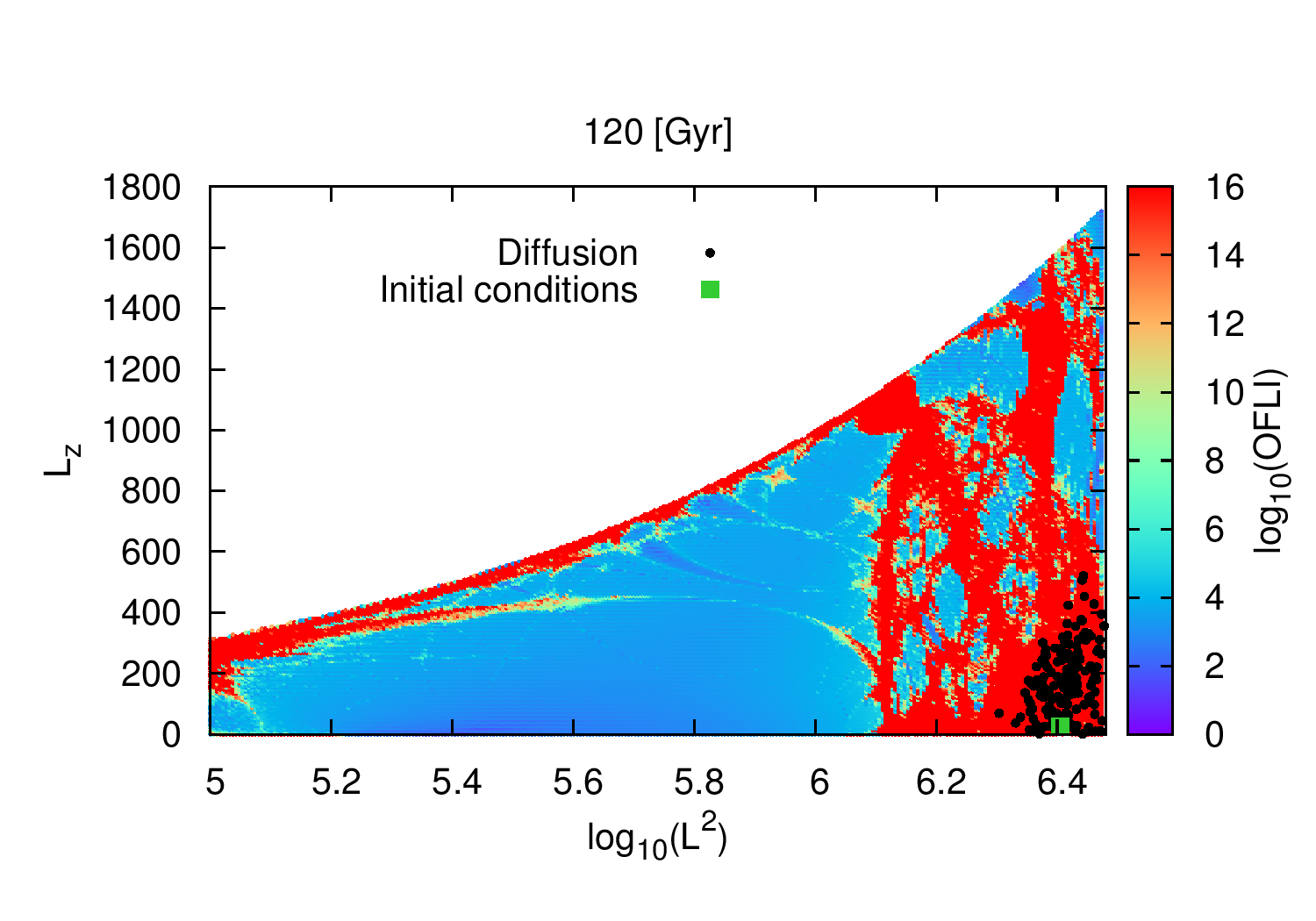}&
\includegraphics[width=0.5\linewidth]{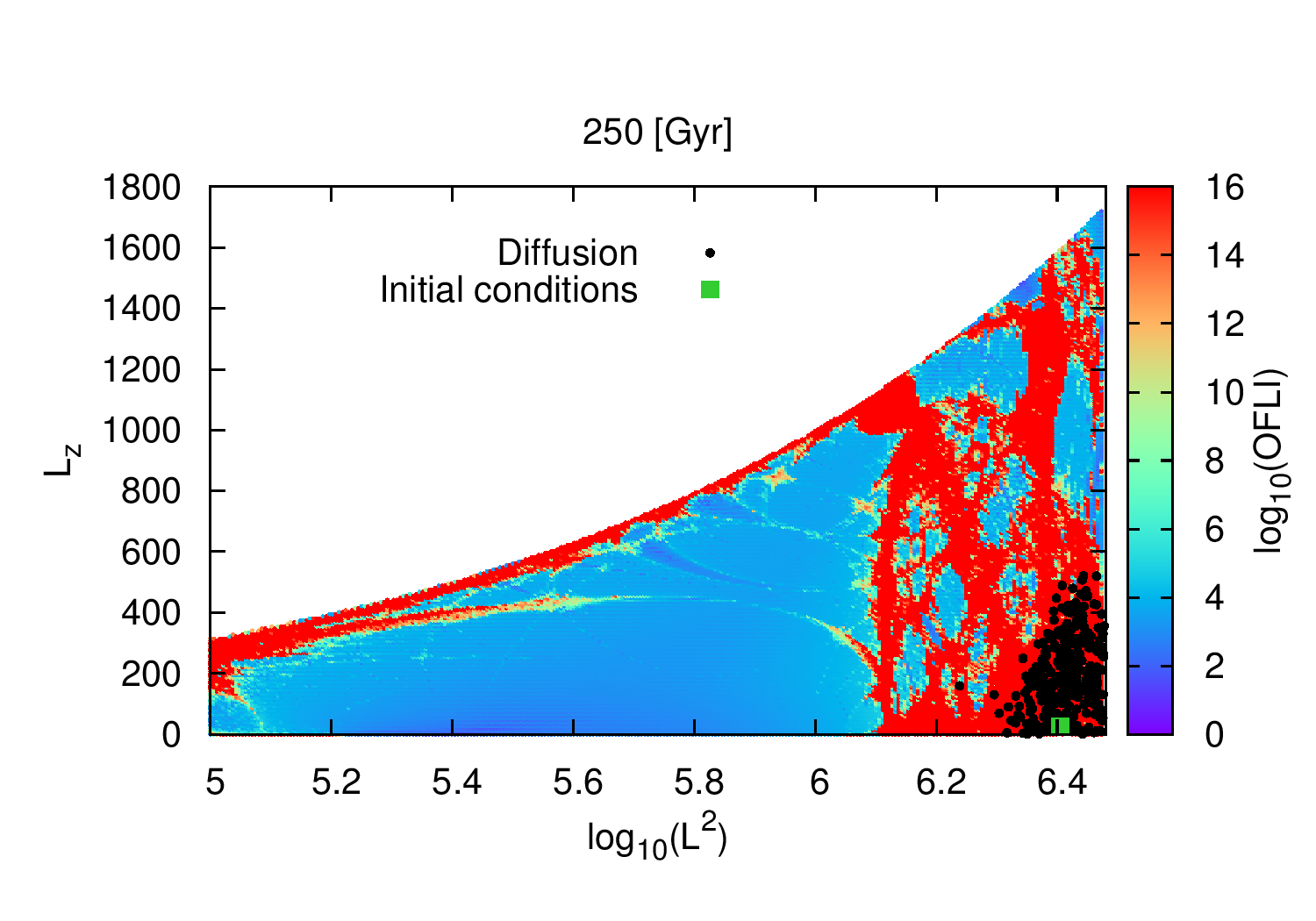}
\end{tabular}
\end{center}
\caption{Long--term diffusion over $120$ Gyr (left panel) and $250$ Gyr (right panel) for ensemble (iv) of initial conditions (depicted in green) overplotted on the Arnold web as in previous figures. Diffusion sweeps a bounded fraction of
the prime integral space --however, the chaotic component is far from being fully connected.}
\label{fig:409}
\end{figure*}

Fig.~\ref{fig:409} also illustrates that the orbits with initial conditions in ensemble (iv) sweep a bounded fraction of prime integral space after $120$ (left panel) and $250$ Gyr (right panel). However, the chaotic component is far from being fully connected, even for this large timescale, since some chaotic domains still remain unexplored. 

From all these experiments we clearly see that in this Hamiltonian representation of the Aq--A2
DM halo, diffusion or chaotic mixing
is completely irrelevant on any realistic timescale.

\section{Discussion and conclusions}
\label{sec:discussion}

The phase space distribution of halo stars in the Neighbourhood of the Sun potentially holds an invaluable source of information about the assembly history of the Milky Way. Stellar streams are the most direct signals of Galactic accretion and their identification in the Solar Neighbourhood is of fundamental importance to the study of galactic dynamics. Our capability to detect these stellar streams might be threatened by local chaotic mixing processes that can smooth out the phase space distribution function on a very short timescale (Section~\ref{sec:introduction}). In this work we have explored whether chaotic mixing can play an important role in shaping the phase space distribution of orbits local to the present position of the Sun  Our results reinforces the idea that this process is very inefficient within a physically meaningful timescale, even within the Solar Neighbourhood.

The degree of substructure in the Solar Neighbourhood's phase space distribution depends on several factors. First, dynamical timescales in the inner regions of the Galaxy are relatively short. In general, stellar streams within this region are expected to be spatially well--mixed \citep{Gould03}. In addition, due (at least) to the triaxial and cuspy nature of the underlying gravitational potential, a fraction of local streams could be evolving on chaotic orbits. Chaos, in the Lyapunov sense, indicates exponential divergence of initially nearby orbits in phase space. Stream stars with chaotic orbits would experience very rapid mixing, with local spatial densities decreasing at an exponential rate such that their detection at the present day would be unlikely. Furthermore, regions filled with chaotic orbits can foster chaotic diffusion, which erases the `dynamical memory' imprinted in phase space and effectively produces a smooth distribution function (Section~\ref{subsubsec:form}).

We have shown that, even though all of these processes are undoubtedly active in the Solar Neighbourhood, they do not necessarily imply a significantly smooth phase space distribution function. One of the key factors in this discussion is the relevant timescale of the system, which serves as an upper bound for the actual time available to develop chaos. In particular, the (dark matter halo) potential in the inner 20 kpc of the Galaxy is not expected to have evolved significantly within the last $\sim 8$ Gyr \citep[$z=1$, see][]{Wetal11,BH14}. The efficiency of chaos should be evaluated with regard to this timeframe.

To characterise the true impact of chaos in shaping local stellar halo phase space structure, we used fully cosmological $N$--body simulations of the formation of Milky Way--like dark matter haloes, coupled with a semi--analytic model of galaxy formation (Section~\ref{subsec:meth-sim}), to sample the phase space distribution of Solar Neighbourhood--like  volumes  (Section~\ref{subsec:meth-init}). 
We modelled the dark matter halo potential with a triaxial extension of the well--known NFW density profile (Section~\ref{subsec:meth-pot}) and used it to integrate the equations of motion, coupled with the first variational equations, to compute the Orthogonal Fast Lyapunov Indicator, OFLI (Section~\ref{subsec:meth-ind-OFLI}). This chaos indicator allowed us to robustly characterise  the dynamical nature of stellar particles. Orbits were classified into three different components: regular, sticky and chaotic. An important difference between these three families is the rate at which the local (stream) density around a given particle decreases as a function of time. While for regular orbits the local density decreases as a power law, for chaotic orbits it does so at an exponential rate. In between we find the so called sticky orbits. Sticky orbits behave as regular for a given period of time, after which their behaviour becomes chaotic.

In all Solar Neighbourhood--like volumes analysed we find that, even within these strongly triaxial 
potentials, only  $\lesssim 20\%$ of the   
stellar particles reveal their chaotic nature within a Hubble time.  We find that $\sim 30\%$ of orbits can 
be characterised as regular. For the remaining $\sim 50\%$, namely `sticky' orbits, it takes in general much
longer than 10 Gyr to reveal their chaotic behaviour. The fraction of sticky orbits is particularly important
because, for halo stars moving on such orbits, chaotic mixing may not have enough time to operate even though
the orbits themselves have an intrinsically chaotic nature. It is important to mention that, in all cases,  
we have considered a simplified representation of the underlying galactic potentials. 
To explore whether this approximation could affect our results, orbits 
of the stellar particles associated with the Aq--A2 local volume were integrated on potentials with axis ratios extracted from the 
remaining four dark matter haloes. In all cases, we kept the virial mass and concentration parameters fixed to those 
associated with halo Aq--A2. If our results would be highly sensitive to the shape of the potential, this change to the 
triaxiality should significantly increase the fraction of chaotic orbits. In all cases we find that the fraction of orbits 
presenting chaotic behaviour within 10 Gyr is $\lesssim 30$\%. The obtained fractions are very similar to the one obtained with the 
axis ratio extracted self--consistently from the dark matter halo Aq--A2. This indicates that, as long as the main sources of chaos 
are included in the model (i.e., central cusp, triaxial shape and its radial dependence), slight variations of the galactic 
potential should not dramatically alter the global dynamics of the  system.

Our results indicates that chaotic mixing, although non--negligible, is not a significant factor in erasing local signatures of accretion events. This is in agreement with the results of G13, who quantified the number of stellar streams in the same Solar Neighbourhood--like volumes considered in this work. Their results suggest that the strongest limitation on quantifying substructure is mass resolution, rather than diffusion due to chaotic mixing. They found that, in the best--resolved local volumes, the number of identifiable streams ranges from $\approx 300$ to 600, in very good agreement with previous analytic predictions \citep[][]{HW99,HWS03}. It is important to note that the orbital classification presented in G13 most likely overestimates the fraction of stellar particles moving on chaotic orbits. As a consequence of the high but still finite particle resolution of these simulations, G13 were forced to track the time evolution of local (stream) densities within  relatively large spheres ($R \gtrsim 4$ kpc). As shown in this work, such large spheres are likely to encompass stellar particles that exhibit very different dynamical behaviours, and hence do not faithfully represent the time evolution of their corresponding local densities.

Despite the optimistic view described above, some relevant caveats of the present work must be discussed and addressed in follow--up work. First of all, the galactic potentials considered in this work are associated only to the underlying distribution of dark matter. In a more realistic model, this potential should also account for the mass distributions of the galactic disc, bulge and a supermassive central black hole \citep[e.g.][]{Siopis2000,KS03,Vetal10,Vetal12}. Note, however, that previous studies including multi--component galactic potentials have successfully identified large amounts of substructure in Solar Neighbourhood--like volumes \citep[e.g.][]{HZ00,GHBL10,SBW11}. Additionally, studies based on stellar haloes obtained from fully cosmological hydrodynamical simulations find an overall fraction of chaotic orbits that, in all cases, is $\lesssim 20\%$, in agreement with our results \citep{Vetal10,Vetal13}. We have also considered static potentials that are not allowed to evolve as a function of time. While an evolving potential could enhance the efficiency of diffusion in phase space \citep[see for instance][]{P13}, previous attempts to characterise the degree of substructure in local volumes, taking into account the variation of the Galactic potential in a cosmological context, have suggested that this effect may not be significant after all. As previously discussed, at least within its inner regions, the Galactic potential is not expected to have evolved significantly during the last $\sim 8$ Gyr. We will further explore the validity of the assumptions adopted in this work in a follow--up study (Cincotta et al. 2015, in preparation).

\section*{Acknowledgements}

NPM and FAG would like to thank Monica Valluri for the very interesting discussion and useful comments that help to improved this manuscript. NPM, FAG, PMC and CMG are grateful to Carles Sim\'o for his illuminating discussions, comments and suggestions. We are also grateful to Volker Springel, Simon White, and
Carlos Frenk for generously allowing us to use the {\it Aquarius} simulations. Finally, NPM wants to thank D.D. Carpintero, F. Bareilles and E. Su\'arez for their assistance with the LP--VIcode as well as with the pre--processing of the data. NPM, PMC and CMG were supported with grants from the Consejo Nacional de Investigaciones Cient\'ificas y T\'ecnicas (IALP UNLP--CONICET) de la Rep\'ublica Argentina (CCT-–La Plata), and the Universidad Nacional de La Plata (FCAG UNLP). FAG and BWO acknowledge support from the US NSF Office of Cyberinfrastructure by grant PHY--0941373 and by the Michigan State University Institute for Cyber--Enabled Research (iCER). BWO was supported in part by NSF grants PHY--0822648 and 1430152: Physics Frontiers Center/Joint Institute for Nuclear Astrophysics (JINA). APC is supported by a CO- FUND/Durham Junior Research Fellowship under EU grant [267209].

\appendix

\section[]{Exponential divergence and variational Chaos Indicators}
\label{sec:CIsrevisited}

The chaos indicators, which are based on the time evolution of the deviation or tangent vector to the flux describing a given dynamical system, measure the rate of divergence of two initially close solutions. In order to quantify the rate of divergence of such two nearby orbits, let us consider a Hamiltonian system defined on a differentiable manifold, the energy surface in the present case.

If we denote by  $\mathcal{H}({\mathbf{p}},{\mathbf{q}})$
the Hamiltonian with ${\mathbf{p}},\,{\mathbf{q}}\in \mathbb{R}^N$, the energy surface is thus defined by
$\mathcal{M}_h=\{{\mathbf{x}}: {\mathcal{H}}({\mathbf{p}},{\mathbf{q}})=h\}$. Further, on introducing the notation

\begin{displaymath}
{\mathbf{x}}=({\mathbf{p}},{\mathbf{q}})\in\mathcal{M}_h,\qquad \mathbf{f}(\mathbf{x})=(-\partial{\mathcal{H}}/\partial{\mathbf{q}},\ \partial{\mathcal{H}}/\partial{\mathbf{p}})\in\mathcal{M}_h,
\end{displaymath}
the equations of motion can be recast as:
\begin{equation}
\dot{\mathbf{x}}={\mathbf{f}}({\mathbf{x}}),
\label{eq:3}
\end{equation}
so that the first variational equations take the form

\begin{equation}
\dot{\mathbf{w}}=\frac{\partial\mathbf{f}}{\partial\mathbf{x}}\mathbf{w},
\label{eq:3.5}
\end{equation}
where $\mathbf{w}$ is the deviation vector and  
${\partial\mathbf{f}}/{\partial\mathbf{x}}$ denotes the Jacobian matrix of $\mathbf{f}$. 
Let $\gamma(t)$ (an orbit) denote a solution of \eqref{eq:3}
for the initial condition $\mathbf{x}_0\in\mathcal{M}_h$. 
Introducing some norm in $\mathcal{M}_h$,
$\|\cdot\|$, we denote 
\begin{equation*}
\delta^{\gamma}(t)=\frac{\|\mathbf{w}(t)\|}{\|\mathbf{w}_0\|},
\end{equation*}
which characterises the Hamiltonian flow in a small neighbourhood of $\gamma(t)$.
Therefore the mean local rate at which nearby orbits to $\gamma(t)$ diverge is given by the largest Lyapunov Characteristic Exponent (lLCE), 
defined as:
\begin{equation}
\mathrm{lLCE}^{\gamma}=\lim_{t\to\infty}\frac{1}{t}\ln\delta^{\gamma}(t)\equiv
\lim_{t\to\infty}\frac{1}{t}\int_0^{\infty}\frac{\dot{\delta^{\gamma}}}{\delta^{\gamma}}\mathrm{d}t.
\label{eq:4}
\end{equation}
The lLCE allows to determine whether an orbit is regular
or chaotic (that is, stable or unstable in the  Lyapunov sense). Indeed, only when the
$\delta(t)$ increases exponentially fast with time the lLCE would be different
from zero.
The inverse of the lLCE, the Lyapunov time, provides the timescale for the manifestation of the
local instability (that is to say the time needed for two nearby orbits to diverge by a factor of one $e$--folding). Whether $\mathrm{lLCE}^{\gamma}$ is null or positive, $\gamma$ is said to be
regular or chaotic, respectively.

The numerical value of
the lLCE for a large but finite time $T$, is the so--called Lyapunov Indicator (LI), which is the most widely used technique of chaos detection. 
Clearly 
the LI is a finite--time approximation of 
$\mathrm{lLCE}^{\gamma}$
given by Eq.~\eqref{eq:4}. 
Therefore, a given  orbit will be classified as either regular or chaotic whether  $\mathrm{LI}$ converges to zero or to a positive value, respectively. The inverse of the 
 $\mathrm{LI}$ is the finite--time approximation to the above defined Lyapunov time \citep[a detailed discussion on the theory and numerical computation of the Lyapunov Characteristic Exponents and, particularly of the lLCE, can be found in the extensive review of][]{S10}.
 
 \section[]{Sticky orbits}
 \label{sec:sticky}
 
Since sticky orbits are important in the present study, we briefly recap their typical behaviour. The phenomenon of stickiness is clearly seen in near--integrable Hamiltonian systems with low--to--moderate perturbations. In this direction, the KAM theory \citep[for a non--rigorous approach see for instance,][]{Ch79,LL83} ensures that most
of the original tori associated with the integrable system survive in the presence of a sufficiently small perturbation. These are the irrational tori, those that satisfy the so--called Diophantine condition. Meanwhile, tori close to a resonance condition are destroyed, leading to unstable chaotic motion. In systems with two degrees of freedom, the two--dimensional invariant tori
divide the energy surface (of dimension 3). For systems of higher dimensionality the scenario is much more complicated, since, for instance the KAM tori no longer divide the energy
surface and the stickiness phenomena, although still present, requires further explanation. Therefore, chaotic orbits cannot enter a (resonant) stability domain due to the presence of an invariant curve that act as a barrier for chaotic motion and so remain confined to a finite width stochastic layer around the island. By slightly increasing the perturbation strength, invariant curves can be broken down and the barriers become only quasi--barriers to chaotic motion. This is due to the intricate
structure of the former invariant curves \citep[][]{A83,MMP84}, an infinite
set of unconnected infinitesimal parts of the earlier curve, strictly speaking a cantor set or cantori \citep[see also][for a qualitative description]{TAV00}. 
Therefore, chaotic orbits starting in a large chaotic domain could in general avoid the cantori. Meanwhile, a chaotic orbit with initial conditions close enough to the cantori might cross the quasi--barriers of the cantorus and stick to the stability island associated with the resonant orbit. Hence, that chaotic orbit would mimic a regular one in the island. Such a chaotic orbit is called `sticky--chaotic' or simply `sticky'. In general, sticky orbits are trapped within thin chaotic layers and might visit the neighbourhood of different stability islands for a rather long time before they
escape (through the cantori) into the chaotic sea (in the context of Poincar\'e surfaces of section). A sticky orbit therefore looks like a regular orbit for a rather long time, until its chaotic nature is clearly exposed -- in other words its chaos onset time is relatively large. 

\bsp

\label{lastpage}

\end{document}